\documentclass{aastex631}

\usepackage{color}

\shorttitle{AASTeX v6.3.1 Sample article}
\usepackage{float}
\usepackage{booktabs}
\usepackage{subfigure}
\usepackage{graphicx}
\usepackage{footnote}
\usepackage{multirow}
\usepackage{color}
\usepackage{amsmath}
\usepackage{cases}
\usepackage{appendix}
\graphicspath{{./}{figures/}}

\begin{document}

\title{Discovery of diffuse $\gamma$-ray emission in the vicinity of G172.8+1.5: An old supernova remnant with different turbulence properties} 


\author[0009-0003-4873-6770]{Yuan Li}
\affiliation{Tsung-Dao Lee Institute, Shanghai Jiao Tong University, Shanghai 201210, People's Republic of China}
\affiliation{School of Physics and Astronomy, Shanghai Jiao Tong University, Shanghai 200240, PRC}

\author[0000-0001-9745-5738]{Gwenael Giacinti}
\affiliation{Tsung-Dao Lee Institute, Shanghai Jiao Tong University, Shanghai 201210, People's Republic of China}
\affiliation{School of Physics and Astronomy, Shanghai Jiao Tong University, Shanghai 200240, PRC}
\affiliation{Key Laboratory for Particle Physics, Astrophysics and Cosmology (Ministry of Education), Shanghai Key Laboratory for Particle Physics and Cosmology, 800 Dongchuan Road, Shanghai, 200240, People's Republic of China}

\author[0000-0003-1039-9521]{Siming Liu}
\affiliation{School of Physical Science and Technology, Southwest Jiaotong University, Chengdu 610031, PRC}

\email{yuanlss17@sjtu.edu.cn}
\email{gwenael.giacinti@sjtu.edu.cn}
\email{liusm@swjtu.edu.cn}

\begin{abstract}
We report the detection of high-energy $\gamma$-ray emission in the vicinity of G172.8+1.5, which is debated as a Supernova remnant (SNR) or an ionized hydrogen (H$_{\rm{II}}$) region. Using 16-yr Pass 8 data from $\emph{Fermi}$ Large Area Telescope (Fermi-LAT), we found the GeV emission can be described by two extended sources with different photon spectra. Among them, the much more extended $\gamma$-ray source SrcA with a Power-law spectrum is spatially coincident with a giant neutral Hydrogen shell structure and several OB stars inside a huge H$_{\rm{II}}$ region. The softer Log-Parabola spectra $\gamma$-ray source SrcB is spatially coincident with a star-forming region with several OB stars, maser sources and IR sources. Gas observation results display a dense molecular cloud surrounding SrcB in the velocity range of [-25,-5] km s$^{-1}$. The possible origins of $\gamma$-ray emission are discussed, including CRs escaped from SNR shock surface and illuminated nearby MC, the young massive star clusters associated with the star-forming region and a pulsar halo generated by an invisible energetic pulsar inside the SNR. Furthermore, under the hadronic scenario, the measured diffusion coefficient in the compact SrcB region is significantly lower than that of the more extended SrcA. This suppression is likely attributed to cosmic-ray-driven instabilities, which reduce the diffusion efficiency in the vicinity of the source relative to the standard conditions in the interstellar medium (ISM). Future advanced analysis from LHAASO observation results would help distinguish the origins of $\gamma$-ray emission in this region and clarify the nature of this source.  
\end{abstract}


\section{Introduction} \label{sec:intro}

Very high energy (VHE: $\geq$100 GeV) surveys with ground-based $\gamma$-ray detector like LHAASO \citep{2024ApJS..271...25C}, H.E.S.S.\citep{2018A&A...612A...1H} and HAWC \citep{2020ApJ...905...76A} have uncovered a population of $\gamma$-ray sources in the TeV regime and revealed the very energetic cosmic ray (CR) accelerators in our Galaxy, most of them are identified as pulsar wind nebula (PWNe), supernova remnants (SNRs), pulsar halos and young massive star clusters (YSCs). However, a large fraction of the sources remain without a firm association, and the multi-wavelength observations of these sources are crucial for revealing their nature and clarifying the origin of CRs around the knee energy range. SNRs interacting with dense molecular clouds (MCs) are expected to be bright in the $\gamma$-ray band. Effectively, the $\gamma$-ray emissions from such systems have been detected by the Fermi Large Area Telescope (\emph{Fermi}-LAT), including SNR G150.3+4.5 \citep{2024A&A...689A.257L}, W44 \citep{uchiyama2012fermi}, W28 \citep{hanabata2014detailed}, $\gamma$-Cygni \citep{2025A&A...700A.143L} and SNR G15.4+0.1 \citep{2023ApJ...945...21L}. The intense GeV $\gamma$-ray emissions from these SNRs are commonly considered to be from the decay of neutral pions generated in inelastic collisions between accelerated protons and the dense gas, and the derived $\gamma$-ray flux depends on the amount of nuclear CRs released and the diffusion coefficient in the interstellar medium \citep[ISM;][]{gabici2009}. On the other hand, over the last decade, more and more observations support the YSC associated with star-forming regions (SFR) to be an important class of factories of Galactic cosmic rays \citep{2019NatAs...3..561A}. CRs could be efficiently accelerated by strong, fast winds of young massive stars and shocks caused by the core-collapse SNs in clusters.
The acceleration efficiency and maximum energy of particles are expected to be enhanced with respect to the standard values derived for a single SNR shock, due to the SN blast wave interacting with fast stellar winds \citep{2020SSRv..216...42B}. Meanwhile, YSCs typically host dense molecular gas to drive the strong star formation makes the hadronic interaction of accelerated CRs with the surrounding dense gas a natural explanation for their $\gamma$-ray emission.
So far, several such systems, e.g., Cygnus cocoon associated with the compact cluster Cygnus OB2 \citep{2011Sci...334.1103A}, RSGC 1 \citep{2020MNRAS.494.3405S}, Westerlund 1 \citep{2012A&A...537A.114A} and Westerlund 2 \citep{2018A&A...611A..77Y} have been detected with the extended $\gamma$-ray structures from GeV to TeV bands.

G172.8+1.5 (G172 hereafter) was reported as an old SNR candidate produced within a very complicated region with a giant neutral H$_{\rm I}$ structure \citep{2012AJ....143...75K}. In this region, several H$_{\rm II}$ regions have been studied \citep{1978A&A....63..325I}, part of them are known associated with a giant molecular cloud at the distance around 1.8 kpc \citep{1981ApJ...246..394E}, in which active star formation is ongoing and at least 14 embedded star-forming clusters with 3-5 Myr old around these H$_{\rm II}$ regions \citep{2008MNRAS.388..729K,2011MNRAS.416.1522C}, indicating their formation could have been triggered by SN explosions or stellar winds. This conclusion is confirmed by recent work \citep{2022MNRAS.517.4669R,2025ApJ...979..162P}. Furthermore, the measured filamentary structure resembles a bow tie morphology \citep{2007MNRAS.379..289K}, indicating the existence of a shock front. In this case, the ROSAT Survey Diffuse X-ray Background Map \citep{1997ApJ...485..125S} revealed the hot X-ray-emitting gas inside some H$_{\rm II}$ complex, which could comes from the supernova explosion, possibly in a cluster and triggered the formation of these H$_{\rm II}$ region \citep{2012AJ....143...75K}. On the other hand, strong thermal absorption caused by nearby H$_{\rm II}$ regions (e.g., at 10 MHz by \citet{1976MNRAS.177..601C} and 22 MHz by \citet{1999A&AS..137....7R}) complicates the measurement of the total radio flux density \citep{2010A&A...515A..64G} of the potential SNR candidate. In the $\gamma$-ray energy band, several point-like sources are listed in the 4FGL-DR4 catalog \citep{Abdollahi2020a,2023arXiv230712546B} released by Fermi-LAT, while none of them is close or spatially coincident with the position of the LHAASO source (1LHAASO J0534+3533) \citep{2023arXiv230517030C}. This intriguing discrepancy in both the extension and position between the GeV and TeV emissions can be perfectly explained by the model C proposed by \citet{2015MNRAS.447.2224B}, which makes this source a good example to understand the propagation and acceleration mechanism in the blast wave of SNRs, see Section \ref{sec:4.1} for details. 

In this work, we conduct a comprehensive analysis of the GeV $\gamma$-ray emission in the vicinity of the G172 region, utilizing 16 years of data from the \emph{Fermi}-LAT instrument. The photon events are selected in an energy range from 100 MeV to 1 TeV, and the results are presented in Sect. \ref{sec:2}. In Sect. \ref{sec:3}, we presents the gas observation results of $^{12}$CO(J = 1-0). Sect. \ref{sec:4} is dedicated to discussing potentially origins of the $\gamma$-ray emission. Lastly, Sect. \ref{sec:5} provides our conclusions.

\section{\emph{Fermi}-LAT Data Reduction}\label{sec:2}

In this part, the standard LAT analysis software \emph{Fermitools} v2.2.0 and \emph{Fermipy} v1.1.6 \citep{2017ICRC...35..824W} are adopted to quantitatively determine the extension and position of the extended source. The latest Pass 8 data are selected from August 4, 2008 (Mission Elapsed Time 239557418) to August 4, 2024 (Mission Elapsed Time 744465605) to study the GeV emission around the G172 region. We chose the "Source" event class together with instrumental response function ``P8R3$\_$SOURCE'' (evclass=128) and event type FRONT + BACK (evtype=3), with the standard data quality selection criteria $\tt (DATA\_QUAL > 0)  \&\& (LAT\_CONFIG == 1)$. To avoid the Earth's limb contamination, only the events with a zenith angle less than 90$\degr$ are selected. Furthermore, photon events in the range 1 GeV - 1 TeV are used for the spatial analysis, while those between 100 MeV and 1 TeV are employed for a more detailed spectral analysis. The analysis is performed within a $14^{\circ} \times 14^{\circ}$ region of interest (ROI) using the standard LAT ScienceTools. All sources from the incremental version of the fourth Fermi-LAT source catalog (4FGL-DR4; \citep{2020ApJS..247...33A,2023arXiv230712546B}) are included in the binned maximum-likelihood analysis \citep{mattox1996likelihood}. Two diffuse background models (IEM, $\tt gll\_iem\_v07.fits$) and ($\tt iso\_P8R3\_SOURCE\_V3\_v1.txt$ ) are adopted, all sources listed in the 4FGL-DR4 catalog are listed in the background model, and all sources within $20\degr$ from the center of ROI and two diffuse backgrounds are included in the model, which is generated by the script make4FGLxml.py\footnote{\url{ http://fermi.gsfc.nasa.gov/ssc/data/analysis/user}}. The likelihood test statistic (TS) is used to evaluate the significance of the $\gamma$-ray sources, defined as $\rm TS = 2(\ln \mathcal{L}_{1} - \ln \mathcal{L}_{0})$, where $\mathcal{L}_{1}$ and $\mathcal{L}_{0}$ are the maximum likelihood values for models with and without the target source, respectively. To assess spatial extension, we define $\rm TS_{ext} = 2(\ln \mathcal{L}_{\rm ext} - \ln \mathcal{L}_{\rm ps})$, where $\mathcal{L}_{\rm ext}$ and $\mathcal{L}_{\rm ps}$ are the likelihood values for extended and point-like templates. Since the extended model introduces only one additional free parameter, the extension significance can be approximated as $\sqrt{\rm TS_{ext}}$ in units of $\sigma$.

\subsection{Morphological analysis}\label{sec:2.2}
\begin{figure*}
    \centering
    \includegraphics[trim={0 0.cm 0 0}, clip,width=0.325\textwidth]{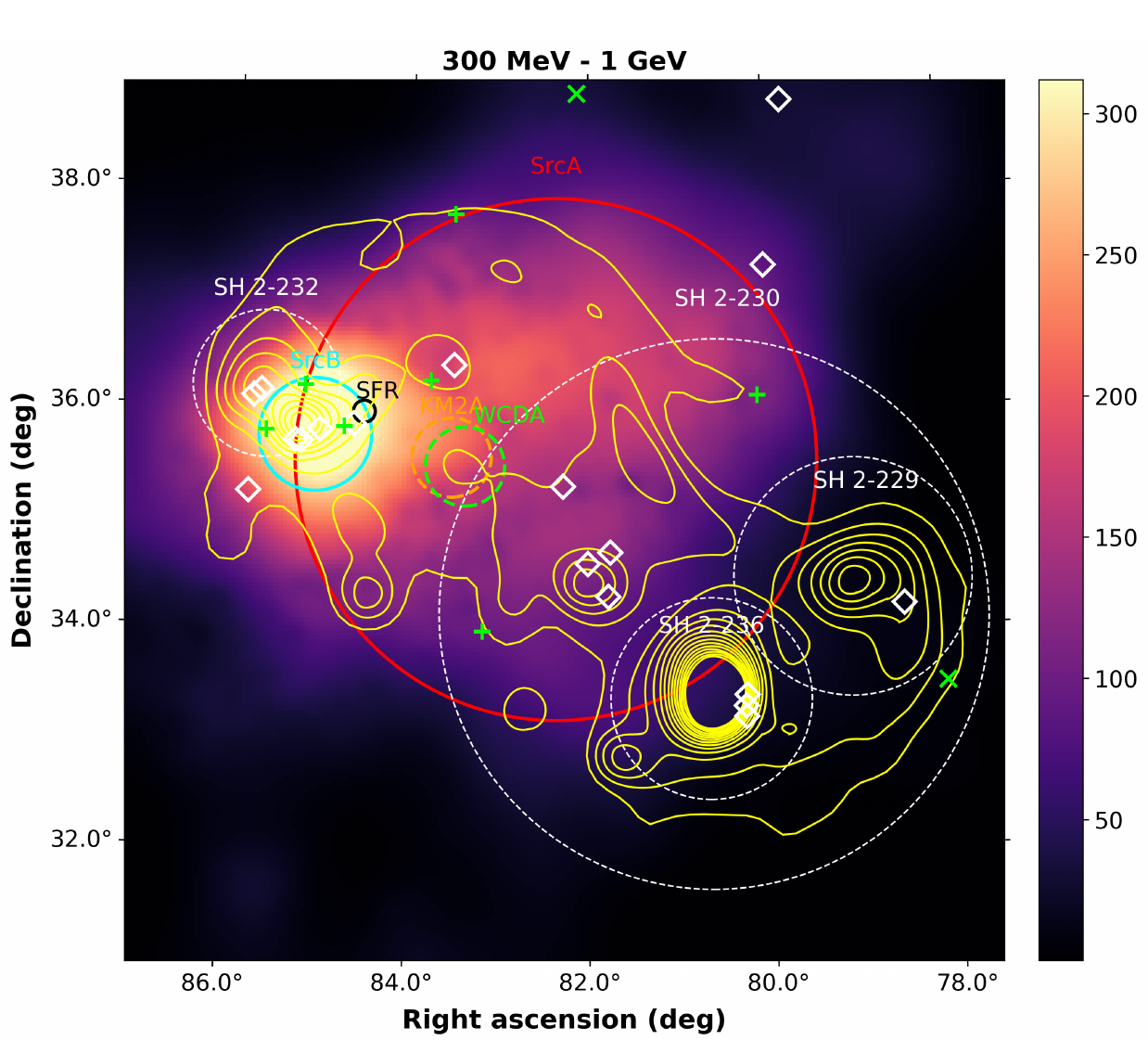} 
    \includegraphics[trim={0 0.cm 0 0}, clip,width=0.325\textwidth]{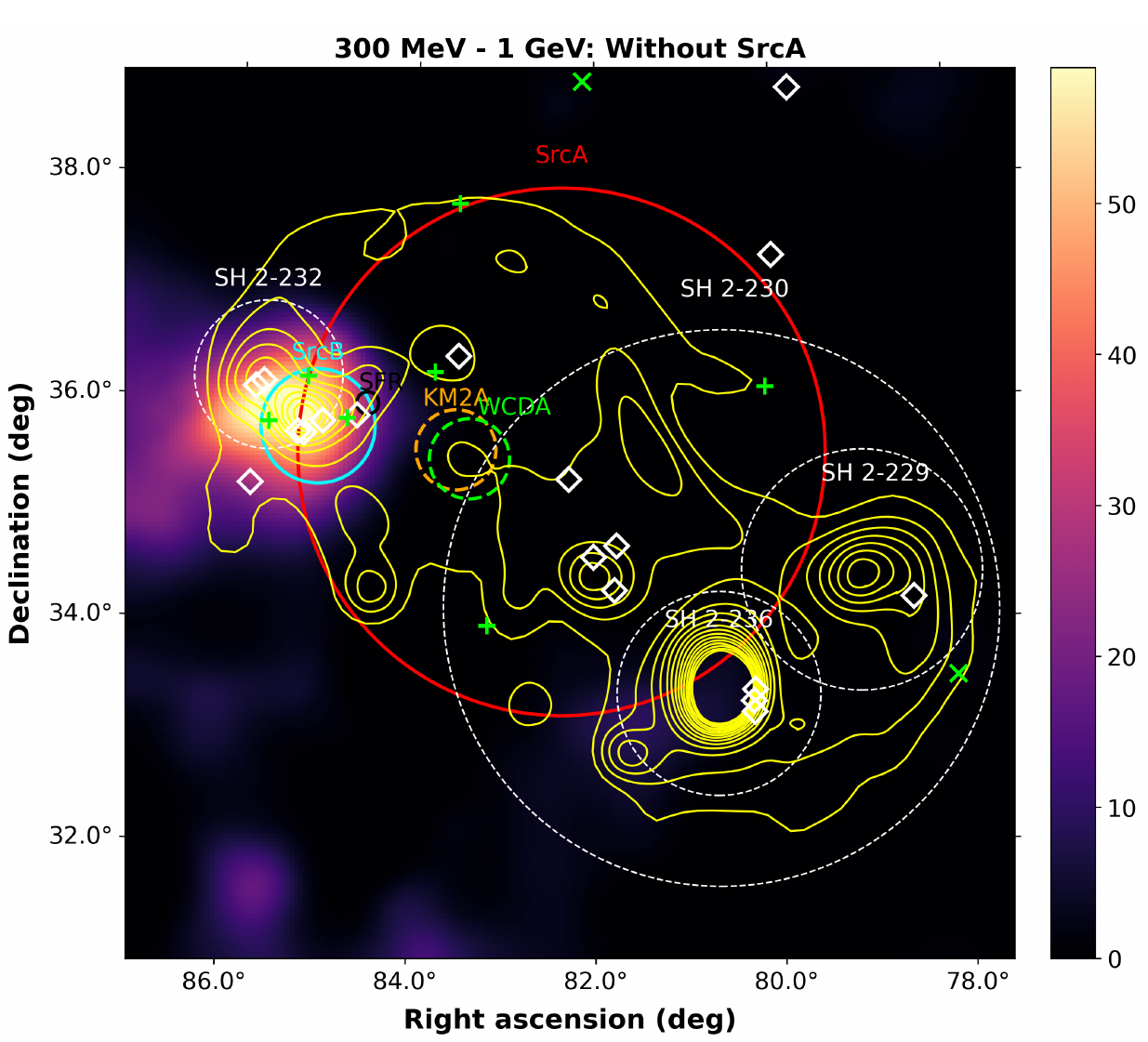} 
    \includegraphics[trim={0 0.cm 0 0}, clip,width=0.325\textwidth]{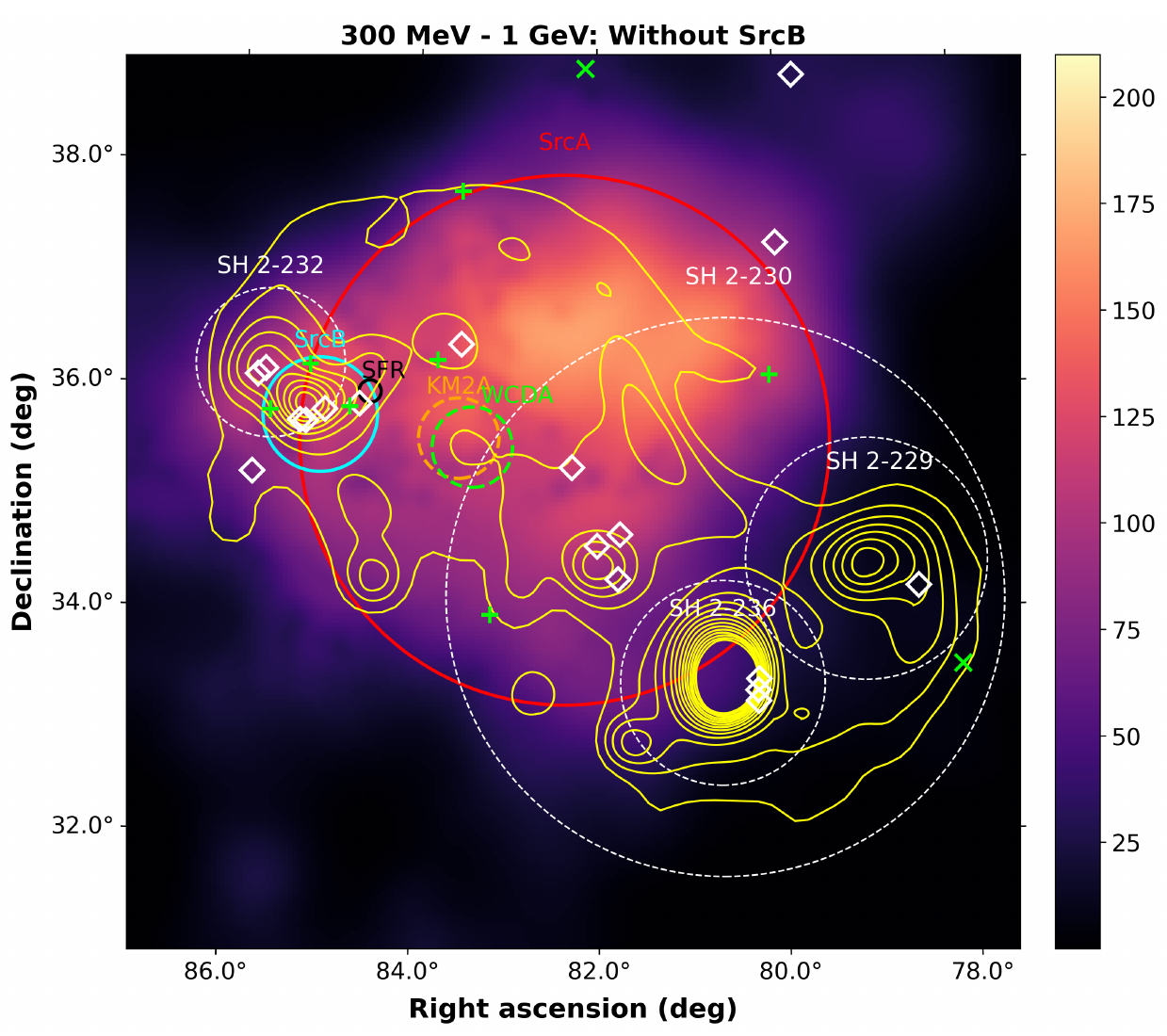} \\
    \includegraphics[trim={0 0.cm 0 0}, clip,width=0.325\textwidth]{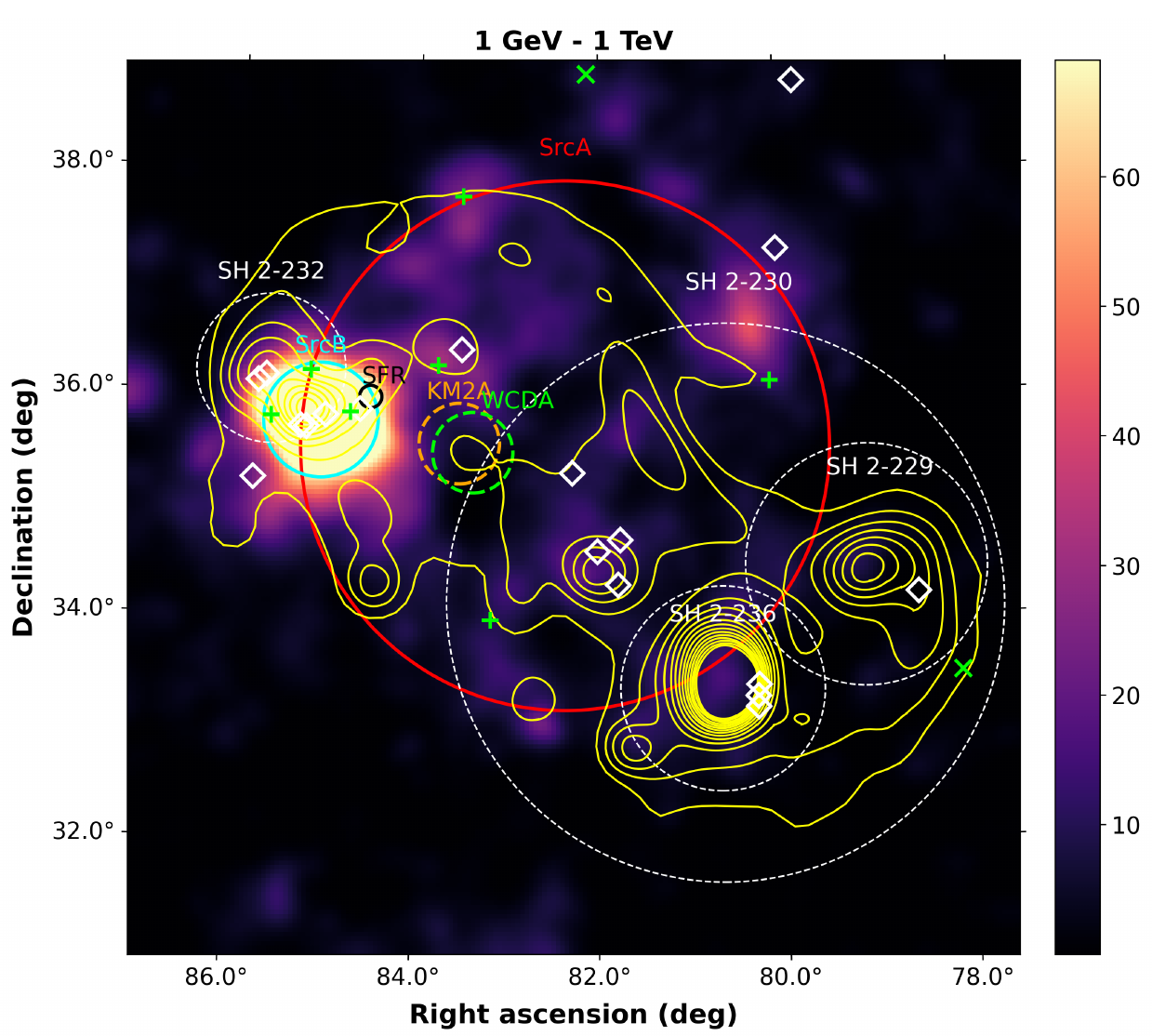} 
    \includegraphics[trim={0 0.cm 0 0}, clip,width=0.325\textwidth]{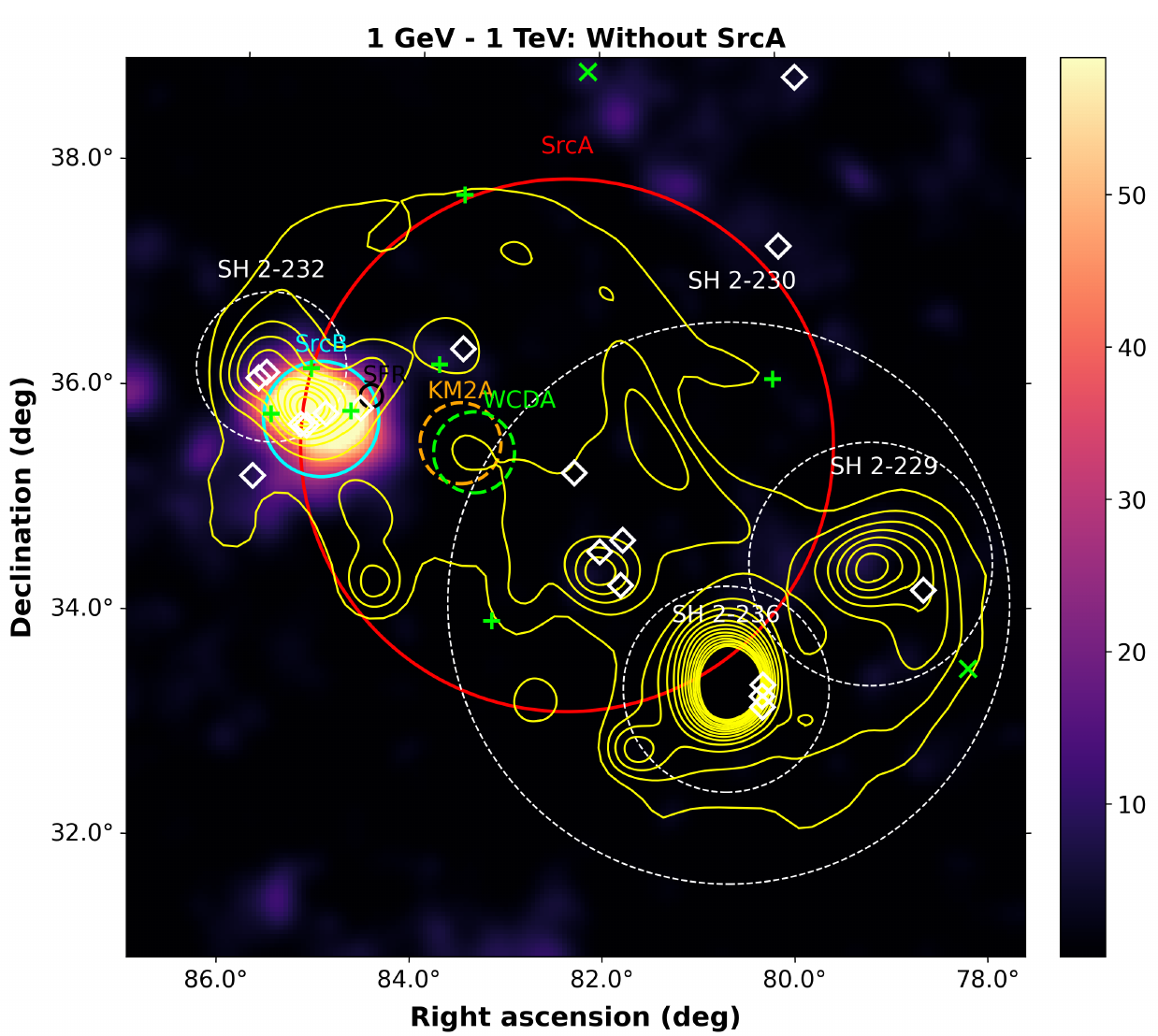} 
    \includegraphics[trim={0 0.cm 0 0}, clip,width=0.325\textwidth]{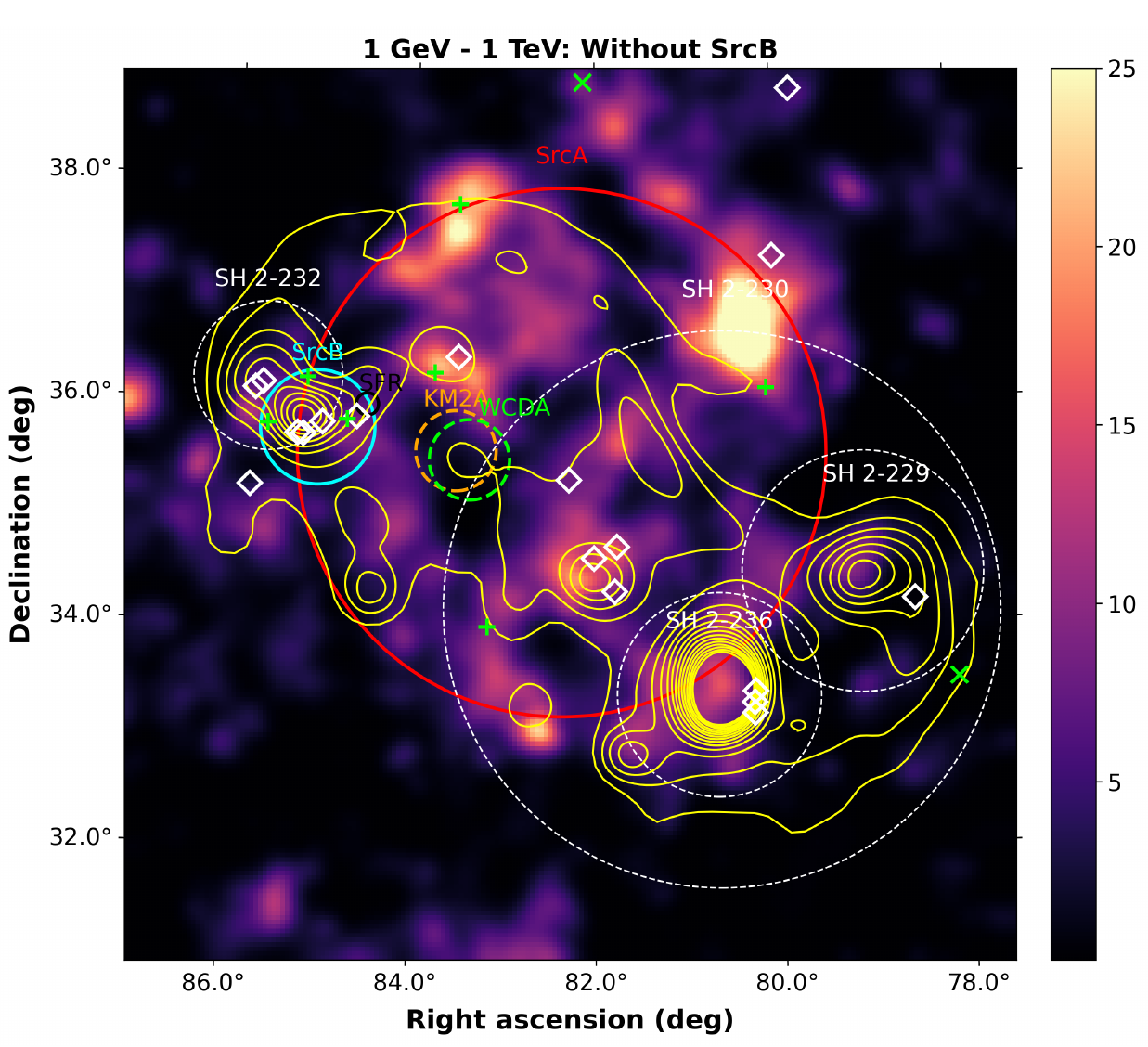} \\
    \caption{TS maps in the vicinity of the G172 region calculated by Fermi-LAT in different energy bands. Top row panels for 300 MeV - 1 GeV energy band, bottom row panels for 1 GeV - 1 TeV energy band. The middle and right column panels show the residual map after subtracting the emission from SrcA and SrcB, respectively. The red and cyan circles show the best-fit R$_{\rm 68}$ extension and position of SrcA and SrcB, respectively. The white dashed circles indicate the position and extension of several ionized hydrogen regions \citep{1976MNRAS.177..601C,2012AJ....143...75K}. The green crosses show the point-like sources listed in 4FGL-DR4 and subtracted in this work. Among them, two green crosses located outside of the SrcA region (red circle) show the point-like sources listed in 4FGL-DR4 and treated as background sources. The white diamond shows the position of several OB stars \citep{2012AJ....143...75K}. The black circle shows the location of a star-forming region G173.185+02.356 \citep{2012A&A...542A...3S}. The green and orange dashed circles represent the 39$\%$ contamination radius r$_{\rm 39}$ of LHAASO J0534+3533 WCDA and KM2A source \citep{2023arXiv230517030C}, respectively. The yellow contour is extracted from Effelsberg 1.4 GHz results \citep{2010A&A...520A..80L}.}
   \label{fig:1}
 \end{figure*}

In the 4FGL-DR4 catalog, there are several point-like sources in the region, we generated TS map by only considering background fitting but not including all point-like sources contribution within the region (shown as green crosses in the Figure \ref{fig:1}), the $\gamma$-ray emission in the 0.3 - 1 GeV and 1 GeV - 1 TeV are shown as the top row and bottom row panels of the Figure \ref{fig:1}, their diffuse emission clearly exhibits spatial inhomogeneity. To assess whether the $\gamma$-ray emission in the region can be modeled as a single extended source, we adopted a uniform disk (Model 2) and a 2D Gaussian (Model 3) and adopted $Fermipy$ to calculate their extension, location and the template performance. Compared with the multiple point-like sources template (Model 1), both the single disk template (Model 2) and the 2D Gaussian template (Model 3) show significantly improved TS value. Compared with the single disk template, the Gaussian template has a marginal TS value improvement with $\sim$7, corresponding to $\sim$ 2.6$\sigma$. After that, we further additionally added another point-like source into Model 2 and Model 3 to test if there is another other source with different spectrum in the region, then we use \emph{gtfindsrc} command to optimized its location and fit it with extended source together, the best-fit value is recorder as Model 4 (Disk + point) and Model 5 (Gaussian + point), respectively, both of them show a non-negligible improvement in TS value. Then we adopted the \emph{Fermipy} package to quantitatively estimate the extension of this new source, the uniform disk and 2D Gaussian results are recorded as Model 6 and Model 7. The calculated results prefer an extended hypothesis by the $\rm TS_{ext} > $16 \citep{2012ApJ...756....5L} in both Model 6 and Model 7. The overall maximum likelihood values and TS values are summarized in Table \ref{tab:1}, and we adopted the Akaike information criterion (AIC;\citep{1974AIC}) to compare the performance between Model 1 and others model, which is defined as AIC = 2k - 2$\ln\mathcal{L}$, and the model with minimum AIC value is preferred, thus the Model 6 is adopted in the following spectra analysis. The emission maps of the two individual components are shown as the middle and right columns for different energy bands in Figure \ref{fig:1}, among them, the much more extended source named as SrcA, and the more compact source named as SrcB.

\begin{table*}
\centering
\caption{\textbf{Spatial templates tested for the GeV $\gamma$-ray emission}}
\begin{tabular}{cccccccc}
\hline \hline
Morphology($>$1GeV)  &R.A., Decl &Best$-$fit Extension (R$_{\rm 68}$)& -log(Likelihood)&Ndf$^{\,\,\text{a}}$&$\Delta${TS}$^{\,\,\text{b}}$&$\Delta${AIC}$^{\,\,\text{b}}$\\
\midrule
Model 1 (4FGL-DR4)&$-$&$-$&56651&28&0&0 \\
\hline
Model 2 (Single Disk)&$85^{\circ}\!.06, 35^{\circ}\!.75$&$0^{\circ}\!.55 \pm0^{\circ}\!.09$&56632&5&38&-84\\
\hline
Model 3 (2D Gaussian)&$85^{\circ}\!.10$, $35^{\circ}\!.78$&$0^{\circ}\!.58 \pm0^{\circ}\!.08$&56629&5&44&-90\\
\hline
\multirow{2}*{Model 4 (Disk + point)}&SrcA: $82^{\circ}\!.56, 35^{\circ}\!.65$&$2^{\circ}\!.36 \pm0^{\circ}\!.14$&\multirow{2}*{56614}&\multirow{2}*{9}&\multirow{2}*{74}&\multirow{2}*{-112}\\
~&SrcB: $-$&$-$&~&~&~\\
\hline
\multirow{2}*{Model 5 (Gaussian + point)}&SrcA: $82^{\circ}\!.75, 35^{\circ}\!.54$&$3^{\circ}\!.25 \pm0^{\circ}\!.23$&\multirow{2}*{56607}&\multirow{2}*{9}&\multirow{2}*{88}&\multirow{2}*{-126}\\
~&SrcB: $-$&$-$&~&~&~\\
\hline
\multirow{2}*{Model 6 (Two Disk)}&SrcA: $82^{\circ}\!.36, 35^{\circ}\!.54$&$2^{\circ}\!.37 \pm0^{\circ}\!.11$&\multirow{2}*{56589}&\multirow{2}*{10}&\multirow{2}*{124}&\multirow{2}*{-160}\\
~&SrcB: $85^{\circ}\!.06, 35^{\circ}\!.75$&$0^{\circ}\!.51 \pm0^{\circ}\!.04$&~&~&~\\
\hline
\multirow{2}*{Model 7 (Two Gaussian)}&SrcA: $82^{\circ}\!.81, 35^{\circ}\!.58$&$3^{\circ}\!.28 \pm0^{\circ}\!.27$&\multirow{2}*{56591}&\multirow{2}*{10}&\multirow{2}*{120}&\multirow{2}*{-156}\\
~&SrcB: $85^{\circ}\!.10, 35^{\circ}\!.78$&$0^{\circ}\!.55 \pm0^{\circ}\!.05$&~&~&~\\
\hline
\hline
\end{tabular}
\label{tab:1}
\\
{{\bf Notes.} $^{(a)}$ Number of degrees of freedom. $^{(b),(c)}$ Calculated with respect to Model 1.}
\end{table*}

\subsection{Spectral analysis}\label{sec:2.2}
\begin{figure*}
    \centering
    \includegraphics[trim={0 0.cm 0 0}, clip,width=0.4\textwidth]{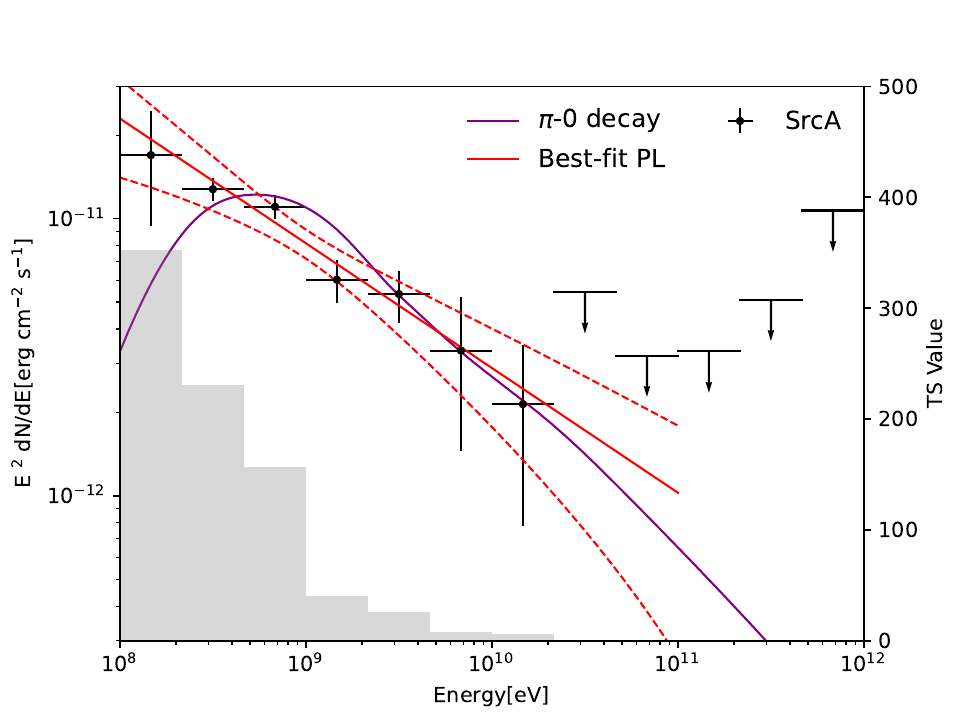}
    \includegraphics[trim={0 0.cm 0 0}, clip,width=0.4\textwidth]{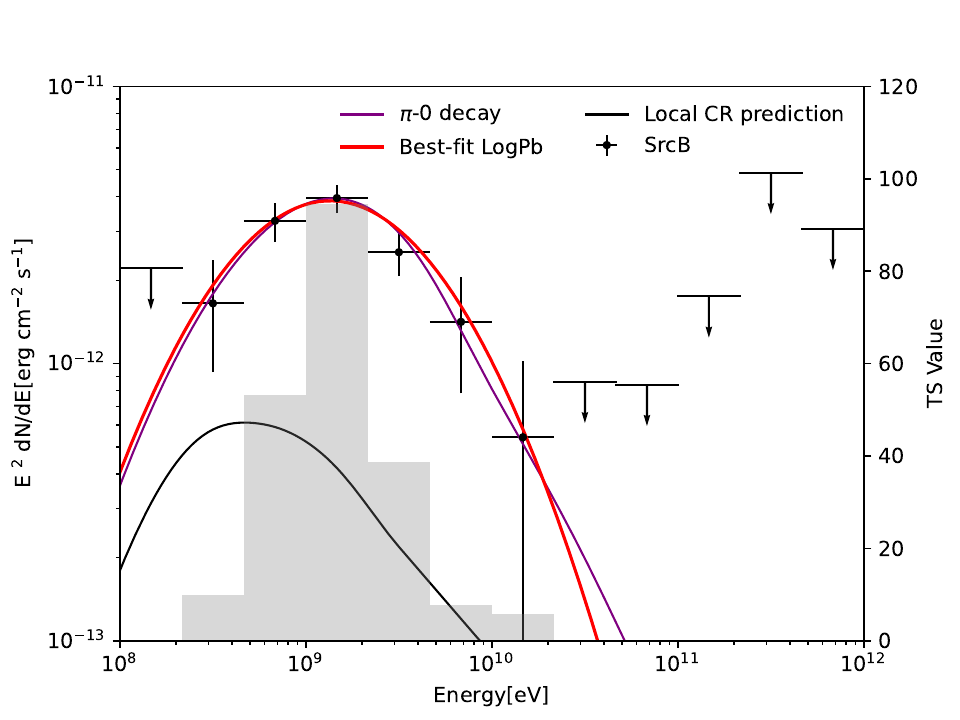}
    \caption{The SEDs of SrcA (left), and SrcB (right). The black data points were derived by Fermi-LAT in the energy range of 100 MeV - 1 TeV. The black arrow indicates the 95$\%$ upper limits and the grey histogram shows the TS value for each energy bin. The red solid and dashed lines show the best-fit PL with 1 $\sigma$ statistical errors for SrcA. The red curve indicates the best-fit LogPb spectrum for SrcB. The solid black line in the right panel represents the predicted local $\gamma$-ray emission assuming that the CR spectra therein are the same as measured locally by AMS-02 \citep{2015PhRvL.115u1101A}. The magenta solid lines represent the hadronic models assuming a power-law and broken power-law proton spectrum for SrcA and SrcB, respectively, details can be seen in Sect.\ref{sec:4.2}.}
   \label{fig:spectra}
 \end{figure*}

Since the best-fit two disk template (Model 6) including SrcA and SrcB was determined, we adopted simple power-law (PL; dN/dE $\propto$ E$^{-\alpha}$), logparabola (LogPb; dN/dE $\propto$ E$^{\rm -(\alpha+\beta log(E/E_{\rm b}))}$) spectra functions to search for the best-fit spectra for each source. For SrcA, the logPb spectral fit results in a significantly lower TS value compared to the single PL model, suggesting that the latter provides a better description of the GeV data points. However, for SrcB, compared with PL assumption, the TS value in LogPb case has a significantly improvement, which can be quantify as $\rm{TS_{curve}}$ and defined as $\rm{TS_{curve}}$=$2(\ln\mathcal{L}_{\rm BPL}-\ln\mathcal{L}_{\rm PL})$\citep{2020ApJS..247...33A}. The obtained value of $\rm{TS_{curve}}$ $\sim$ 128 corresponds to a significance level of $\sim$ 11.3 $\sigma$ with only one additional free parameter, thus we conclude that there is an energy spectral break ($\rm{E_{b}}$) at $\sim$ 2.45 GeV in the SrcB spectrum, and LogPb has better performance. Additionally, we separated the events in the 100 MeV–1 TeV energy range into twelve logarithmically equal intervals and performed the same likelihood fitting analysis in each interval to obtain the spectral energy distributions (SEDs) for each source. All sources' normalizations are left free, but the energy break and spectral indices are fixed to their best-fit values. Using a Bayesian approach, we provide upper limits with a 95\% confidence level for bins with TS values below 5.0 \citep{helene1983}. Figure \ref{fig:spectra} displays the plotted best-fit global spectra and the corresponding SEDs. The measured results are summarized in Table \ref{tab:2}.

\begin{table}[!htb]
\centering
\caption{Results of spectral analysis in the energy range of 100 MeV - 1 TeV}
\begin{tabular}{ccccccc}
\hline\hline
SrcA & SrcB  & DoF & -log(Likelihood) &$\Delta${TS}$^{\,\,\text{a}}$\\
\hline
PL    & PL    & 4 & -3190827 &  0          \\
LogPb & PL    & 5 & -3190806 &  -42         \\
LogPb & LogPb & 6 & -3190870 &  87         \\
PL    & LogPb & 5 & -3190891 &  128         \\      
\hline
\end{tabular}
\\
{{\bf Notes.} $^{(a)}$ Calculated with respect to the all power-law case.}
\label{tab:2}
\end{table}

\begin{table}[!htb]
\centering
\caption{Parameters of the best-fit spectral models in the energy range of 100 MeV - 1 TeV}
\begin{tabular}{cccccccc}
\hline\hline
Source & $\Gamma_{1}$& $\Gamma_{2}$ & $\rm{E_{b}}$ & Photon flux \\
~      & $\alpha$            &     $\beta$   &  (GeV) & (ph cm$^{-2}$ s$^{-1}$) \\
\hline
SrcA   & 2.45 $\pm$ 0.16 & $-$  & $-$  & (1.24 $\pm$ 0.04)$\times$10$^{-7}$ \\
SrcB &0.87 $\pm$ 0.18 & 0.33 $\pm$ 0.06  & 2.43   & (1.02 $\pm$ 0.15)$\times$10$^{-8}$ \\
\hline
\end{tabular}
\label{tab:spectra}
\end{table}

\section{Gas observations}\label{sec:3}
In this section, we utilize the CO composite survey data \citep{2001ApJ...547..792D} to trace the H$_{\rm 2}$. We assume that the CO (J=1–0) line intensity at 115 GHz (2.6 mm) acts as a linear tracer of the H$_{\rm 2}$ column density. The column density of H$_{\rm 2}$ in this area can be calculate by using conversion factor $X_\mathrm{CO}=2\times10^{20} \ \rm{cm^{-2} \ K^{-1} \ km^{-1} \ s}$ \citep{bolatto2013}, then the column density $N_\mathrm{H_2}$ is calculated as $N_\mathrm{H_2} = X_\mathrm{CO} \times W_\mathrm{CO}$, thus the mass of the molecular complex can be derived from the $W_\mathrm{CO}$:
\begin{equation}\label{eq:massco}
 M={\mu m_\mathrm{H}} D^2 \Delta\Omega_\mathrm{px} X_\mathrm{CO} {\sum_\mathrm{px}} W_\mathrm{CO} \propto N_\mathrm{H_2},
\end{equation}

\begin{figure*}
     \centering
    \includegraphics[trim={0 0.cm 0 0}, clip,width=0.4\textwidth]{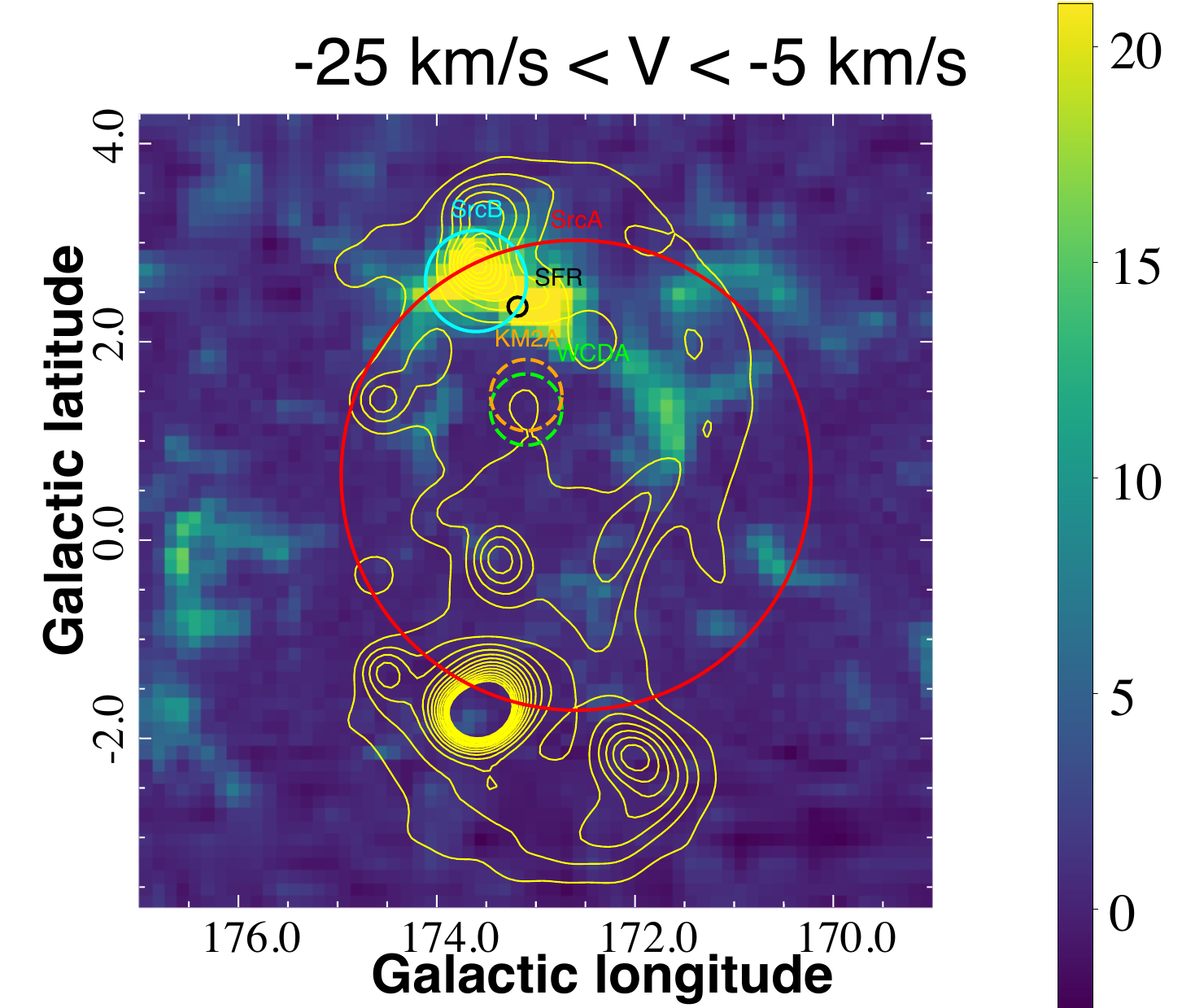}
    \includegraphics[trim={0 0.cm 0 0}, clip,width=0.49\textwidth]{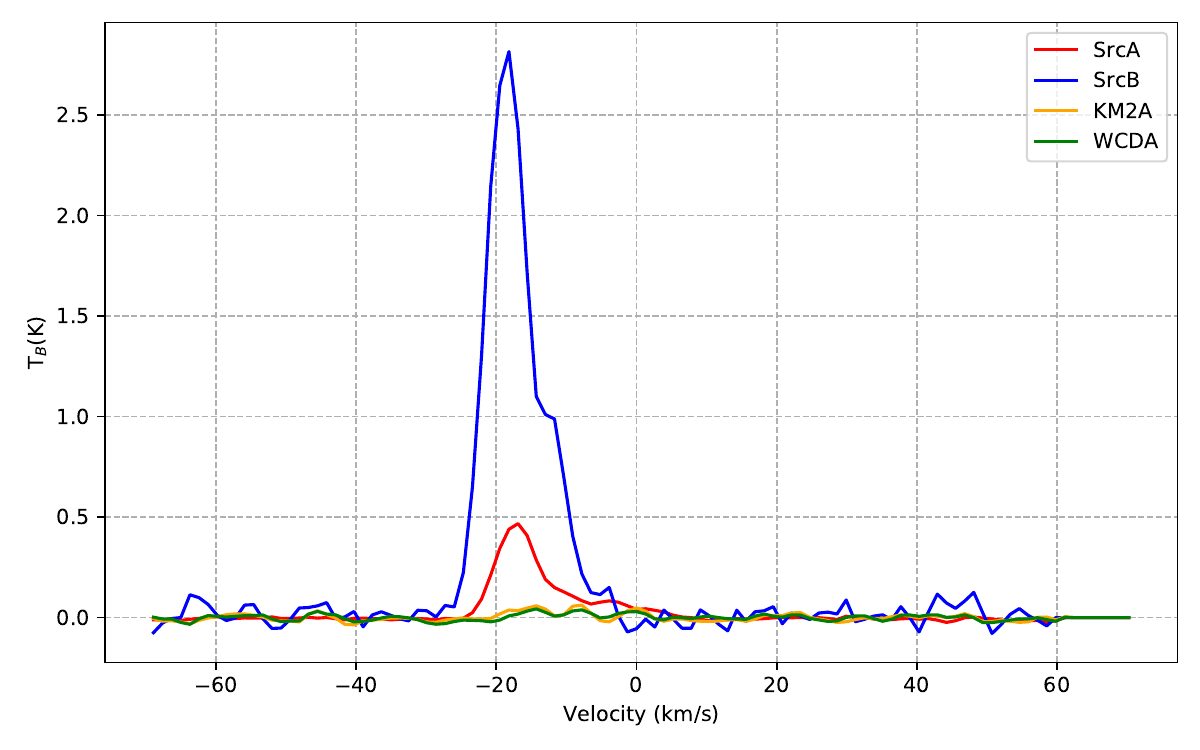}
    \caption{Left: Integrated $^{12}$CO(J = 1-0) emission intensity (K km s$^{-1}$) toward G172 in the velocity range of [-25,-5] km s$^{-1}$. All labels are kept the same as Figure \ref{fig:1}. Right: $^{12}$CO(J = 1-0) spectra of gas inside SrcA, SrcB and LHAASO source regions, respectively.}
\label{fig:4}
\end{figure*}

With $\mu$ set to 2.8 in this expression, the relative helium abundance is 25$\%$, $m_\mathrm{H}$ represents the mass of a hydrogen nucleon. Each pixel's subtended solid angle is determined by $\Delta\Omega_\mathrm{px}$. The velocity binning of the data cube is taken into consideration by the term ${\sum_\mathrm{px}} W_\mathrm{CO}$. It is computed by scaling by the velocity bin size after adding up the map content for the pixels in the target sky region and the specified velocity range. The calculated results from the left panel of Figure \ref{fig:4} suggest there is good spatial coincidence between SrcB and gas distribution. 

On another hand, previous studies \citep{2014ApJS..212....1A} have shown that H$_{\rm II}$ regions and expanding shells can substantially distort the local velocity field, and the CO emission might exhibits multiple velocity components, deviating from predictions of standard rotation models \citep{2014ApJ...783..130R}, leads the calculation of kinematic distances in this region may not be reliable \citep{2009ApJ...700..137R}. Considering the high velocity H$_{\rm I}$ emission features are confined inside the radio continuum filaments associated with the large ionized hydrogen complex \citep{2012AJ....143...75K}, here we directly adopted the distance of $d$ = 1.8 kpc suggested by \citet{2012AJ....143...75K}, which derived from a systemic velocity of -20 km s$^{-1}$ for the H$_{\rm II}$ complex. In this case, the physical extension for SrcA and SrcB can be calculated by R = $d \times$ $\theta_i$ ($i$=A, B). By adopting $\theta_B$ = 0.51$\degr$, the total gas mass within SrcB is estimated to be about $\rm{M_B} = 6.12\times 10^{5}\ d_{1.8}^{2} \ M_{\odot}$. Assuming a spherical geometry of the gas distribution, we estimate the volume to be $\rm{V_B} = {{4\pi \over3}R^3}$, the average $\rm H_2$ cubic density in this region is about $\rm{n_{B}} = \rm 1445\ d_{1.8}^{-1} \ cm^{-3}$. Similarly, by using the $\theta_{Atotal}$ = 2.37 $\degr$, the total gas mass within SrcA region can be calculated as $\rm{M_{Atotal}} = 3.06\times 10^{6}\ d_{1.8}^{2} \ M_{\odot}$, corresponding to $\rm{n_{{Atotal}}}$ = $\rm 72\ d_{1.8}^{-1} \ cm^{-3}$. Then the gas mass between sphere A and sphere B can be calculated as $\rm{M_{A}} = \rm{M_{{Atotal}}} - \rm{M_B}$ = 2.45 $\times 10^{6}\ \rm{d_{1.8}^{2}} \ \rm{M_{\odot}}$, corresponding to $\rm{n_{{A}}}$ = $\rm 57\ d_{1.8}^{-1} \ cm^{-3}$. The gas density measured within LHAASO source region is $\rm{n_{L}} = \rm 3.3\ d_{1.8}^{-1} \ cm^{-3}$. The spectra of CO (J=1–0) emission are extracted within several regions, shown as the red and blue solid curve in the right panel of Figure \ref{fig:4}, which displays significantly peaks for both SrcA and SrcB region in the velocity intervals of $\sim$ [-25,-5] km $s^{-1}$. In contrast, the CO spectra within WCDA and KM2A regions are flat. 

\section{Discussion of the possible origins of the $\gamma$-ray emission}\label{sec:4}
\subsection{CRs escaped from SNR and illuminated nearby gas}\label{sec:4.1}

It is possible that part of the observed $\gamma$-ray excess could originate from imperfect modeling of the Galactic diffuse $\gamma$-ray background, particularly, the H$_{\rm II}$ gas component is not included in the $Fermi$-LAT diffuse background templates \citep{2016ApJS..223...26A}. Nevertheless, as illustrated by the black solid line in Figure \ref{fig:spectra}, the predicted $\gamma$-ray flux from H$_{\rm II}$ gas assuming the CR spectra are the same as those measured locally \citep{2015PhRvL.115u1101A} is far below the observed $\gamma$-ray emission. Therefore, the $\gamma$-ray flux detected from SrcB cannot be explained by uncertainties in the diffuse background modeling.

Given the good spatial correlation between the MC and SrcB, its $\gamma$-ray emission is likely produced through hadronic $\pi^0$-decay. In this scenario, CRs that escaped from the SNR shock front interact with the nearby MC gas, giving rise to the observed $\gamma$-ray signal. Under such an assumption, the injected proton spectra for SrcA and SrcB should be identical. The corresponding escape spectra in Zone A and Zone B, which account for the $\gamma$-ray emission from SrcA and SrcB, respectively, can be used to constrain the model parameters such as the injection energy, diffusion coefficient, and turbulence type. Here we adopt a scenario in which protons were injected instantaneously into two uniformly emitting regions (Zone A and Zone B) about 330,000 years ago, corresponding to the estimated age of the SNR \citep{2012AJ....143...75K}. The injected proton spectrum is assumed to follow a power law with an exponential cutoff:
\begin{eqnarray}
Q(E) = {Q_0} E^{-\Gamma} \exp \left(- \frac{E}{E_{\rm p, cut}} \right).
\label{eq:p_spectra}
\end{eqnarray}

\begin{figure*}
    \centering
    \includegraphics[trim={0 0.cm 0 0}, clip, width=0.45\textwidth]{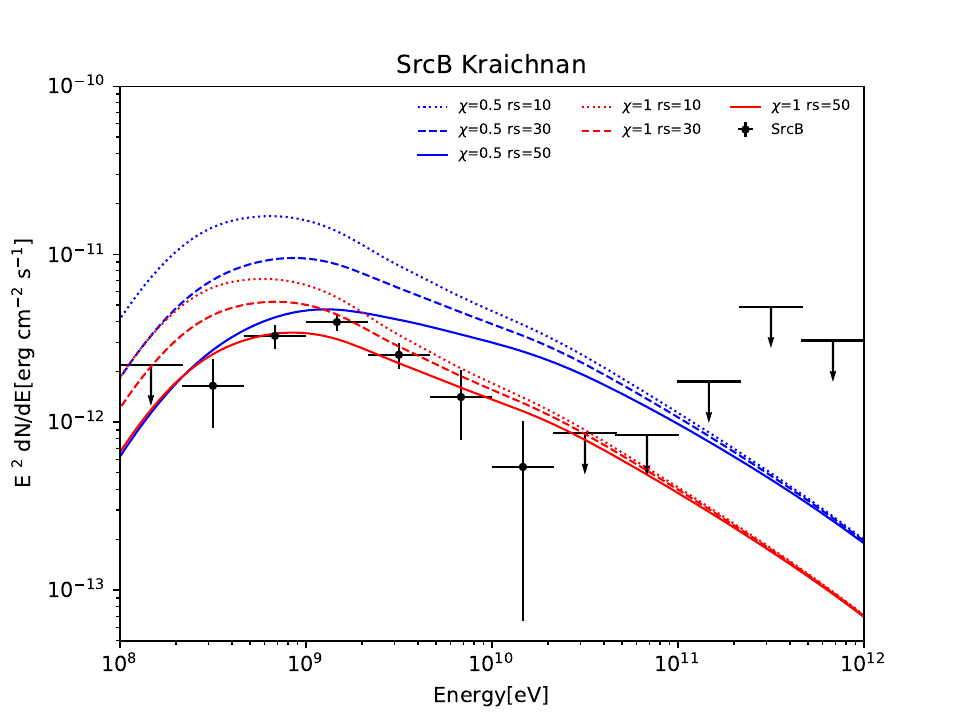}
    \includegraphics[trim={0 0.cm 0 0}, clip, width=0.45\textwidth]{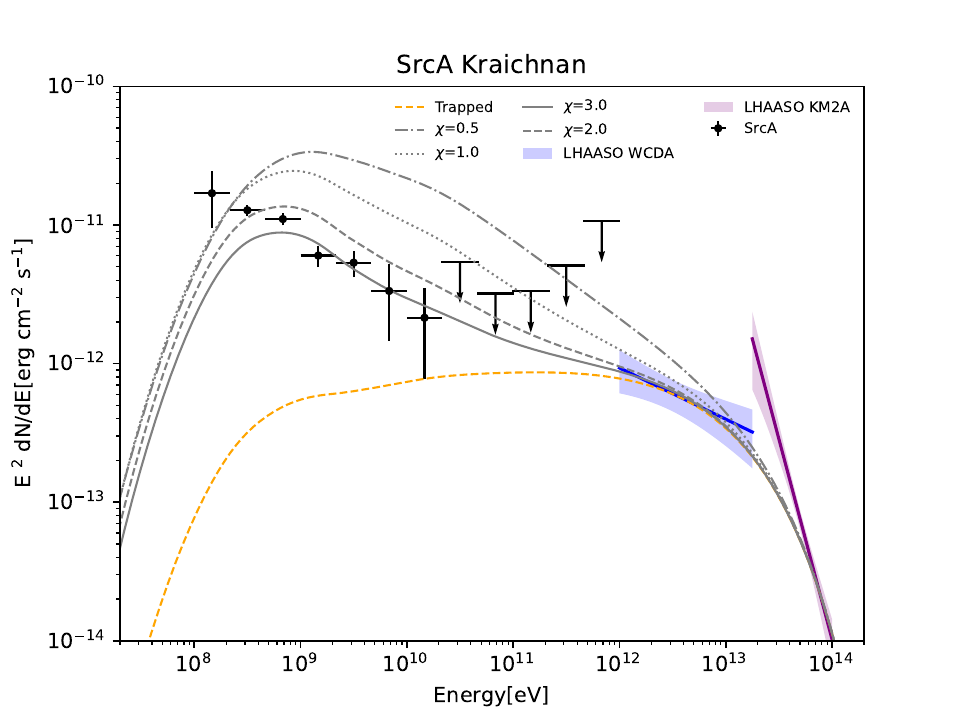}\\
    \includegraphics[trim={0 0.cm 0 0}, clip, width=0.45\textwidth]{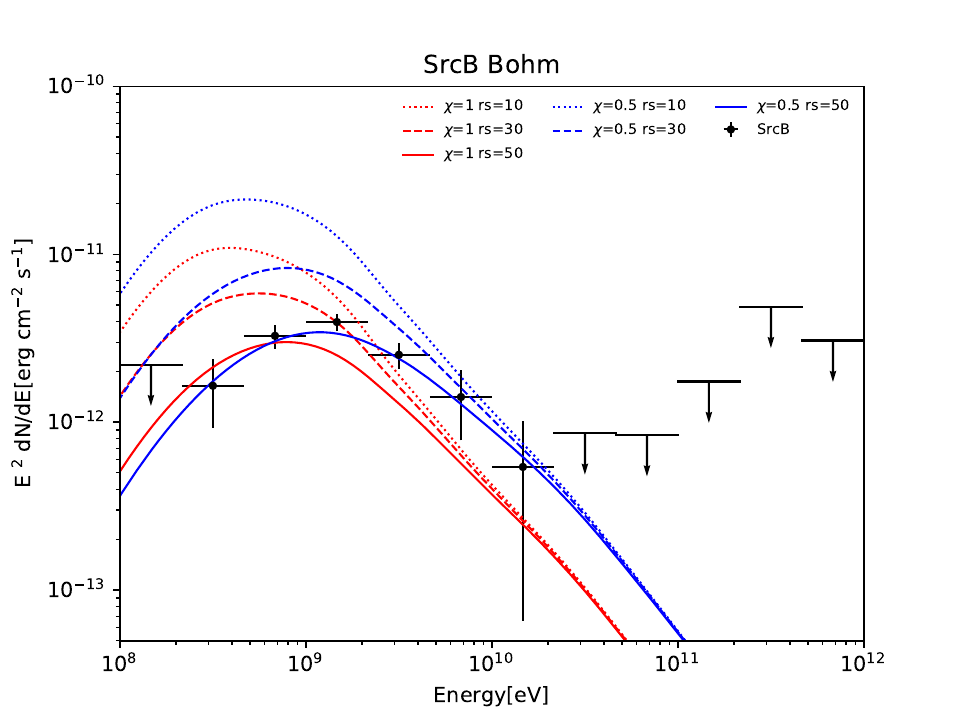}
    \includegraphics[trim={0 0.cm 0 0}, clip, width=0.45\textwidth]{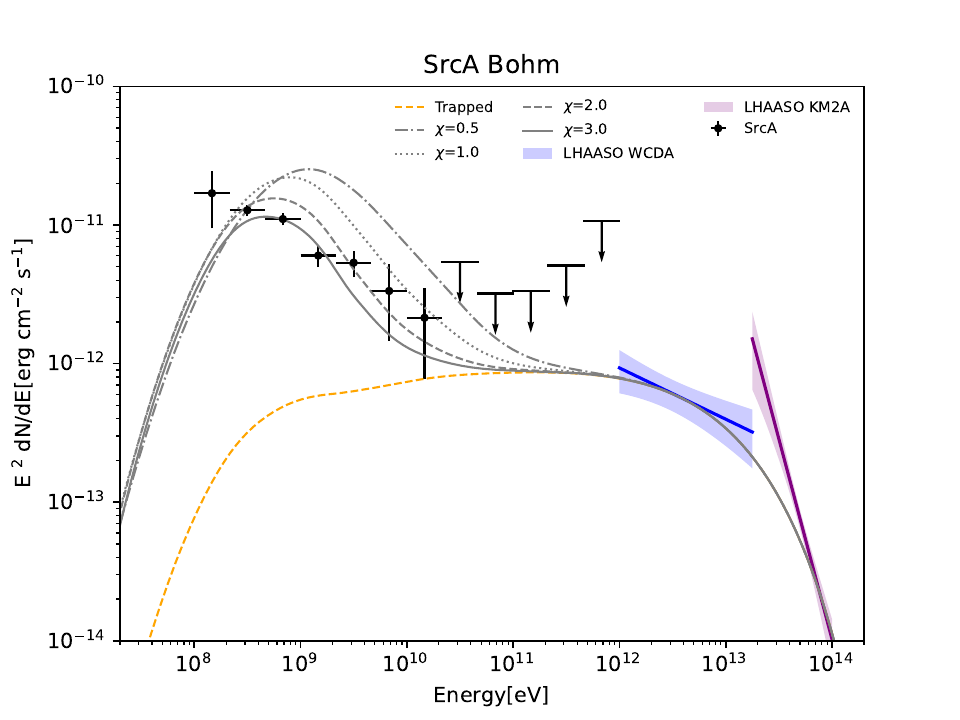}
    \caption{Hadronic modeling of multi-wavelength $\gamma$-ray spectra of SrcA (right column) and SrcB (left column), respectively. The solid, dashed, dotted and dash-dotted lines show different r$_{\rm s}$ value influence towards fitted results. The orange dashed line corresponds to the trapped component with a fixed value of 2.0. The Blue and purple butterflies are extracted from the 1LHAASO J0534+3533 WCDA and KM2A spectra results \citep{2024ApJS..271...25C}. The blue and red  fit lines in the left column panels correspond to the contribution from escaped ions under different r$_{\rm s}$ and $\chi$ value. The gray fit lines in the right column panels show the total contributions from trapped and escaped ions under different $\chi$ value assumptions.}
    \label{fig:5}
\end{figure*}

In our model, $\Gamma$ and $E_{\rm p,cut}$ denote the spectral index and the cutoff energy of the protons, respectively. For simplicity, we assume an injected proton spectrum with $\Gamma=2.0$, consistent with the typical prediction of diffusive shock acceleration, and fix the cutoff energy at $E_{\rm p,cut}=105$ TeV. The resulting hadronic $\gamma$-ray emission is displayed as the orange dashed curves in the right column panels of Figure \ref{fig:5}. The proton energy above 1 GeV for the fit line is estimated to be $W_{\rm p,L}=3.82\times10^{48}(n_{\rm L}/3.3{\rm cm^{-3}})^{-1}$ erg. The distribution of escaped protons within the emission zone is then derived by following the method described in \citet{2012MNRAS.419..624T,2020ApJ...897L..34L}:
\begin{equation}
    N_p(E,t)=\frac{Q(E)}{[4 \pi D(E) T]^\frac{3}{2}}  \exp\left(\frac{-r_{\rm s}^2}{4 D(E) T}\right) \label{equation:3}
\end{equation}

In this formula, the uniform diffusion coefficient can be described as $D(E) = \chi D_0 (E/E_0)^\delta$ for $E > E_0$, where $D_0 = 1 \times 10^{28}$ cm$^2$ s$^{-1}$ at $E_0 = 10$ GeV, $\delta = 1/2$ for Kraichnan turbulence and $\delta = 1.0$ for Bohm diffusion \citep{2006ApJ...642..902P,2013A&ARv..21...70B}, respectively. Due to line-of-sight projection effects, the real distance between the molecular complex and the SNR cannot be precisely determined (see, e.g., HESS J1912+101; \citealp{2023ApJ...953..100L}). To account for this uncertainty, we treat $r_{\rm s}$ as a free parameter, which represents the effective distance between the CR injection site and the illuminated molecular cloud. Given an injected proton spectrum $Q(E)\propto E^{-\Gamma}$ and the above diffusion formula, the escaped proton distribution $N_p(E)$ develops a low-energy break at $E_{p,\rm break}$ defined by the condition $\sqrt{4D(E_b)T}\simeq r_{\rm s}$. Above this break, the spectral shape steepens to $N_p(E) \propto E^{-\left(\Gamma+\frac{3}{2}\delta\right)}$. The total injected energy in protons is parameterized as $W_{\rm inj}=\eta E_{\rm SN}$, where $\eta\approx0.1$ denotes the efficiency of converting the kinetic energy of the supernova explosion into CR protons, and $E_{\rm SN}$ is taken to be the canonical $10^{51}$ erg \citep{2013A&ARv..21...70B}. The expected $\gamma$-ray emission from the interaction between these protons and the ambient gas is then computed with the $\emph{naima}$ package \citep{zabalza2015naima}.

\begin{figure}
     \centering
    \includegraphics[trim={0 0.cm 0 0}, clip,width=0.5\textwidth]{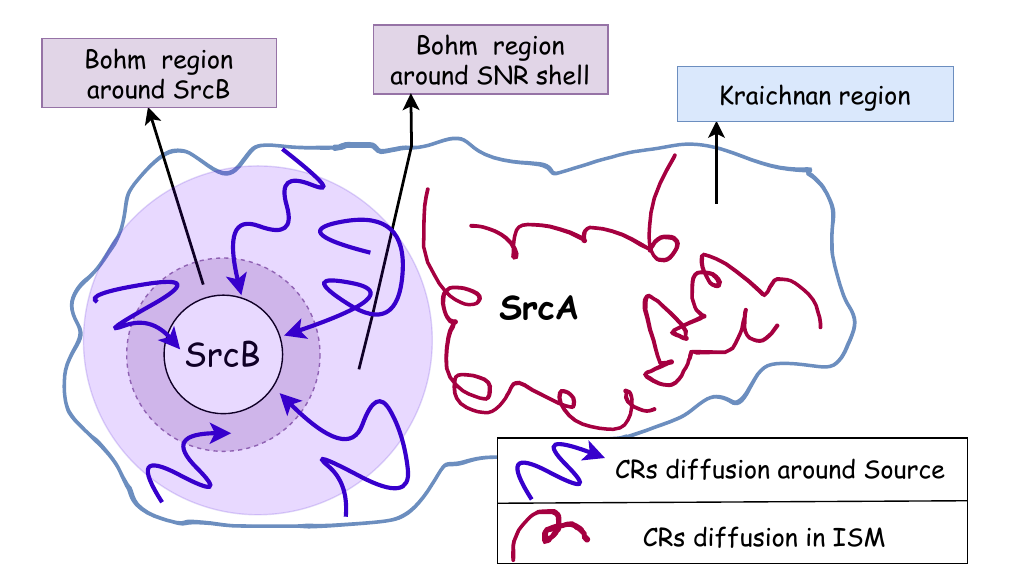}
    \caption{Sketch of the CR propagation in different emission zones.}
\label{fig:6}
\end{figure}

For SrcB, in the left column panels of Figure \ref{fig:5}, the calculated escaped ions contribution have the same amplitude with the observed GeV data points from SrcB, indicating the diffusion coefficient around SrcB region ($\chi$=0.5 to $\chi$=1.0) is much lower than the standard Galactic diffusion coefficient ($\chi$=3.0) \citep{2013A&ARv..21...70B}. Especially, in the Kraichnan scenario, the spectra index in the higher energy band would be harder than observation results, which makes the Bohm diffusion more plausible, and this phenomenon matches the prediction from previous simulation results, suggesting a much slower diffusion process within the Bohm region \citep{1949bohm,2004MNRAS.353..550B}. Considering the different types of $r_{\rm s}$ fit lines, the real physical distance between MC and SNR shock surface could be constrained to around 50 pc. For the much more extended $\gamma$-ray source SrcA, we adopted $r_{\rm s}$ = 50 pc into the calculation, the gray fit lines shown the total contribution from escaped ions and trapped ions with different $\chi$ value show as the right column panels of Figure \ref{fig:5}, indicating the diffusion coefficient within SrcA region can be constrained around $\chi$=2.0 to $\chi$=3.0, which is several times larger than the value predicted within SrcB region. However, based on the fit lines shown in the right column panels, it is difficult to rule out Kraichnan turbulence or Bohm diffusion, both of which show good performance with the measured GeV data points. This result suggests that the turbulence within the SrcA region may be more complex, likely a mixture of Kraichnan turbulence and Bohm diffusion. Such a scenario is actually more physically reasonable, the sketch of the physical model is illustrated in Figure \ref{fig:6}: the SrcB region and parts of the surroundings of the SNR shell are dominated by Bohm diffusion, while in the regions farther from the shell, the $\gamma$-ray emission is dominated by Kraichnan-type turbulence in the ISM. This conclusion is consistent with the prediction from numerical simulations \citep{2004MNRAS.353..550B,2014ApJ...783...91C}.

Furthermore, since the 1LHAASO J0534+3533 WCDA and KM2A are suggested as a point-like source in \citet{2024ApJS..271...25C}, and the 95$\%$ statistic uncertainty is suggested to be 0.18$\degr$, which is much smaller than the 2.37$\degr$ radius extension size measured in GeV band, makes the scenario unutterable if the $\gamma$-ray emission originates from hadronic process and the SNR G172 is the counterpart of LHAASO source. However, based on the Model presented in \citet{2015MNRAS.447.2224B}, the higher energy particles can be trapped in a relatively small region compared with the lower energy particles. In this scenario, these high-energy CRs correspond to the CRs that were accelerated early during the expansion phase of the SNR with high shock velocity, which were advected inside the SNR and remain trapped in tangled magnetic fields from CR-driven instabilities. More specifically, from Figure 5 of \citet{2015MNRAS.447.2224B}, the CRs with energies $\gtrsim$ 250 TeV (corresponding to the $\sim$ 25 TeV photons in KM2A energy band) are predicted to be trapped in a small region around the center of SNR, whose radius is $\sim$ 16$\%$ of the SNR shock radius, which is $\sim$ 2.0 degree as visible in Figure \ref{fig:1} and Figure \ref{fig:4}, therefore this model predicts that the high-energy CRs should be trapped in a region with $\sim$ 0.32$\degr$ radius. This is consistent with the measurement from LHAASO that 1LHAASO J0534+3533 has a 39$\%$ contamination radius of 0.36$\degr$. Correspondingly, the escaped low-energy CRs from the SNR illuminate nearby MC, generating SrcA and SrcB. Their spectra can be fitted by distinct diffusion coefficient parameters, implying that the CRs propagated through different environments before reaching the MCs. Crucially, the fitting results of the spectral radius parameter $r_{\rm s}$ suggest that the SNR shock front and the MC are physically separated by a distance of $\sim$ 50 pc. This result indicates that the apparent spatial overlap between the SNR shell and MC distribution is affected by the projection effect, and the observed gamma-ray emissions from SrcA and SrcB are produced by CRs escaping the SNR and bombarding the MC at different locations, rather than by a direct shock-MC interaction.

\subsection{YSC associated with SFR}\label{sec:4.2}
YSCs associated with star-forming regions (SFRs) can also contribute to the population of Galactic CRs \citep{2019NatAs...3..561A}. Considering SrcB is spatially coincidence with the ionized region SH 2-232 \citep{2012AJ....143...75K} shown in Figure \ref{fig:1}, which is suggested to be associated with a giant MC with 1.8 kpc distance (the velocity range $\sim$ -20 km s$^{-1}$)\citep{1981ApJ...246..394E,1982ApJS...49..183B}. Also, several bright OB stars, infrared sources and maser sources are found with the detected star-forming region G173.185+02.356 \citep{1998A&AS..132..211H,2011MNRAS.414.1526D,2012A&A...542A...3S}, which is reported that embedded in several tens of star clusters \citep{2003A&A...397..177B,2008MNRAS.388..729K,2011MNRAS.416.1522C}(e.g. CBB1, CBB2, [SUH2003] 71, [SUH2003] 72, etc.). Moreover, the open clusters [BDS2003] 73, [KPS2012] MWSC 0629, [FSR2007] 0787 are also reported \citep{2013A&A...558A..53K,2013MNRAS.436.1465B}. Furthermore, two infrared bubbles ([HKS2019] E68 and [HKS2019] E69) \citep{2019PASJ...71....6H} indicate the potential particle acceleration sites and the possibility of production of Galactic CRs \citep{1979ApJ...231...95M,1983SSRv...36..173C}. These proofs support the plausibility that the $\gamma$-ray emission originates from the YSCs associated with the SFR. In this scenario, we also assume that the $\gamma$-rays are produced in the pion-decay process from the interaction of the CRs with nearby gas. Considering there is no energy break for SrcA spectra and a significant energy break for SrcB spectra, here we adopted a simple power-law for SrcA and a broken power-law for SrcB, respectively. The latter formula can be seen in some $\gamma$-ray massive SFR examples \citep{2019ApJ...881...94L,2022MNRAS.517.5121G}: 
\begin{equation}
\resizebox{0.6\hsize}{!}{$
\frac{dN_{\rm p}}{dE} \propto  
\begin{cases}
\left(\frac{E}{E_0}\right)^{-\alpha_{p1}} \qquad \qquad \qquad \qquad;  E < E_{\rm e,break}  \\
\left(\frac{E_{\rm p,break}}{E_0}\right)^{\alpha_{p2}-\alpha_{p1}} \left(\frac{E}{E_0}\right)^{-\alpha_{p2}} \,\,\,\,; E \geq E_{\rm p,break}
\end{cases}
$}
\end{equation}
\\
Here, we set $E_{\rm p,break}$, $\alpha_{p1}$ and $\alpha_{p2}$ as free parameters for the fitting, the gas density for SrcB is adopted from Section \ref{sec:3}, and the fit results are shown as the magenta lines in Figure \ref{fig:spectra}. The derived best-fit spectra index for SrcA is $\sim$2.7 and the total energy above 1 GeV are calculated as W$_{\rm p,A}$ = 3.65$\times$ 10$^{48}$ (n$_{\rm {A}}/57 \ \rm cm^{-3}$)$^{-1}$ erg. Similar, the derived best-fit parameters are $\alpha_{p1}$ = 0.21, $\alpha_{p2}$ = 3.48 and $E_{\rm p,break}$= 13.0 GeV for SrcB. The total energy above 1 GeV is calculated as W$_{\rm p,B}$ = 2.97$\times$ 10$^{46}$ (n$_{\rm {B}}/1445 \ \rm cm^{-3}$)$^{-1}$ erg.

Based on Figure \ref{fig:1}, there are several OB stars located in the area, and the number of O stars over a dozen \citep{1982ApJ...263..777G,2004ApJS..151..103M}. By adopting a typical stellar wind of 2000 km/s $\sim$ 3000 km/s, the average wind kinetic power of $L_{\rm {wind}}$ is suggested to be $1\times 10^{36}$ erg/s $\sim$ $1\times 10^{37}$ erg/s \citep{1975ApJ...200L.107C,1982ApJ...259..282A}, and the total kinetic wind power should be around $1\times 10^{37}$ erg/s $\sim$ $1\times 10^{38}$ erg/s, which is close to the kinetic luminosity of Cygnus Cocoon with $2\times 10^{38}$ erg/s \citep{2019NatAs...3..561A}. Furthermore, since the embedded star-forming clusters are around 3-5 Myr old around these {H$_{\rm II}$} regions, by adopting the average 4 Myr old for the cluster, the diffusion coefficient can be calculated as $\rm{D = r^2 / 4T}$ \citep{2019NatAs...3..561A}. By adopting the SFR distance d = 1.8 kpc, $\theta_A$ = 2.37$\degr$, $\theta_B$ = 0.51$\degr$, the radius for SrcA and SrcB can be calculated as 74 pc and 16 pc, respectively. Then the diffusion coefficient inside Zone A and Zone B is calculated as $\sim 1.5 \times 10^{26} \rm{cm^2 s^{-1}}$ and $\sim 4.8 \times 10^{24} \rm{cm^2 s^{-1}}$. We notice that the diffusion coefficient for Bohm diffusion is $\rm{D_{B}} = c\times r_{\rm g} /3$ = 3.3 $\times 10^{24} \rm(E/ 1 TeV) \rm{(10\mu G/B) cm^2 s^{-1}}$, where r$_{\rm g}$ is the gyro radius. Thus for the 100 GeV protons responsible for 10 GeV $\gamma$-rays, the Bohm diffusion coefficient can be calculated $\sim$ 3 $\times 10^{23} \rm {cm^2 s^{-1}}$ when interstellar magnetic field B $\sim$ 10 $\mu$G is adopted for SrcA region. These results suggested that the diffusion coefficient inside the SrcB region may be only one order of magnitude higher than the Bohm diffusion limit, and is significantly lower than the standard value in the interstellar medium. 

\begin{figure*}
    \centering
    \includegraphics[trim={0 0.cm 0 0}, clip, width=0.4\textwidth]{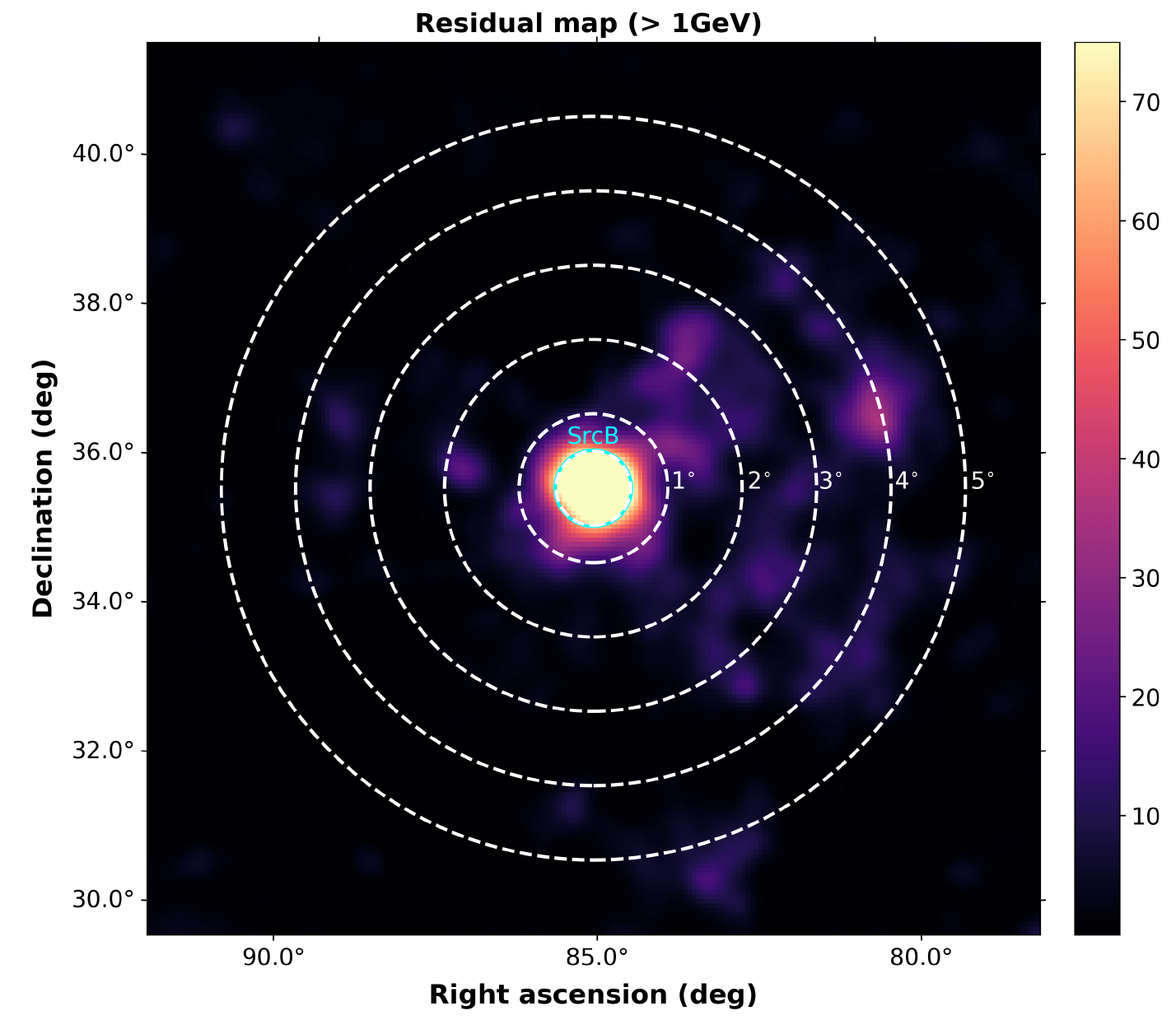}
    \includegraphics[trim={0 0.cm 0 0}, clip, width=0.45\textwidth]{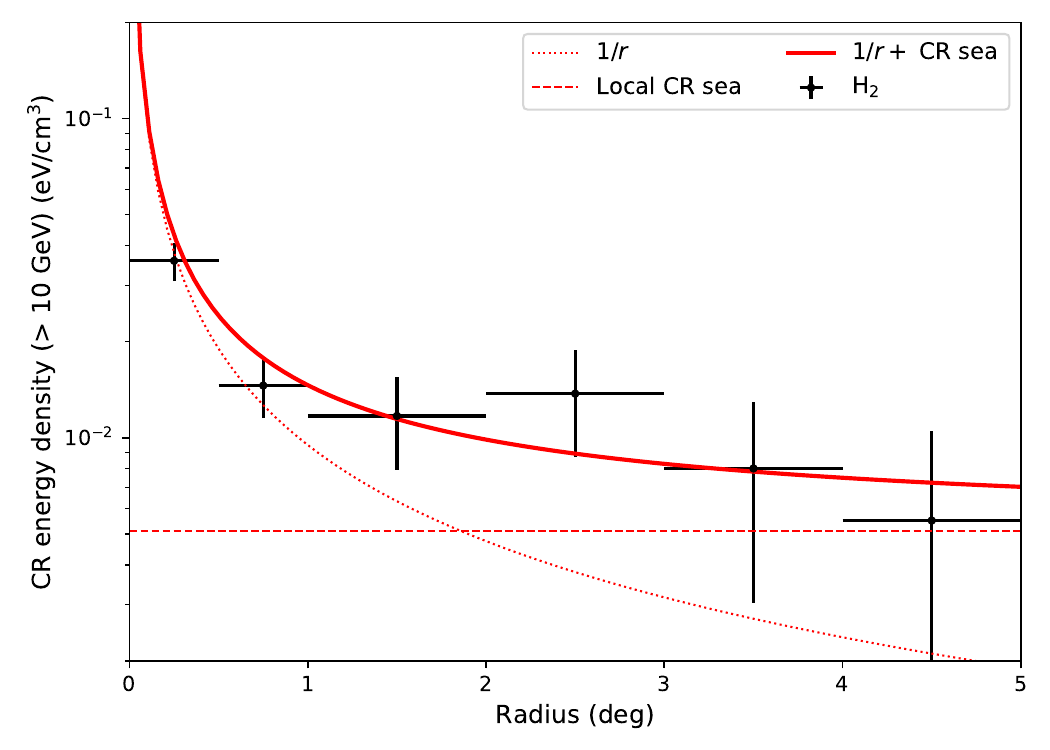}
    \caption{Left panel: TS residual map above 1 GeV overlaid with the rings used to derive the CR energy density. Right panel: Calculated CR density profile around the G172 region, the data points are the $\gamma$-ray emission above 1 GeV of G172. The red solid curve is estimated from adding the local CR sea (dashed line) contribution and the 1/r distribution (dotted line) together.}
    \label{fig:7}
\end{figure*}

Furthermore, if the young massive star clusters are indeed accelerated CRs, then the CRs will inevitably escape and their distribution can be derived from the $\gamma$-ray emission and gas distributions. For example, Cygnus Cocoon \citep{2019NatAs...3..561A}, Westerlund 1 \citep{2023A&A...671A...4H} and Westerlund 2 \citep{2018A&A...611A..77Y} all display a 1/r CR profile, indicating the continuous injection and diffusion-dominated propagation of CRs from the massive star clusters. For SrcB, if the distance of 1.8 kpc is adopted, the physical extension would only be $\sim$ 16 pc, while if we assume that the $\gamma$-ray emission in SrcA also comes from CR released by YSC located inside SrcB and propagation in the ISM, then the physical extension size $\sim$ 80 pc, which is also close to the limitation of known $\gamma$-ray YSC examples. Here, we divided the 5$\degr$ radius region from the SrcB center into 1 disk and 5 rings, with radii of [0:0.5], [0.5:1], [1:2], [2:3], [3:4], [4:5] degrees. The residual TS maps in these regions are shown in the left panel of Figure \ref{fig:7}. By adopted the $\gamma$-ray production cross-section \citep{2014PhRvD..90l3014K}, the CR density profile can be calculated by fitted $\gamma$-ray flux and gas distribution in each region, the method in details can be seen in \citet{2013A&A...560A..64H} and \citet{2020A&A...639A..80S}, the fitting results are shown in the right panel of Figure \ref{fig:7}. 

The above calculated low diffusion coefficient for the SrcB region combined with the relatively high normalization factor of CR density shown in Figure \ref{fig:7} supports the potential Bohm diffusion scenario. This result is also consistent with the prediction in Sect. \ref{sec:4.1} that the SrcB is dominated by Bohm diffusion, and the much more extended SrcA region is dominated by the mixture of Kraichnan turbulence and Bohm diffusion in a different region. Furthermore, recent studies by \citet{2021NatAs...5..465A} and \citet{2022A&A...666A.124A} suggest that flatter spatial profile distributions are more favored over the 1/r profile predicted for continuous injection in models of particle acceleration by stellar wind termination shocks \citep{2021MNRAS.504.6096M}.

\subsection{Leptonic scenario: SNR and pulsar halo}
Since the data points below 1 GeV from SrcA are not well fitted by the hadronic scenario due to the pion bump around 1 GeV, here we further consider a leptonic scenario for the detected $\gamma$-ray emission. Although there is no known energetic pulsar in the region, the observed gamma-ray emission could be explained by a radio pulsar that is invisible in the $\gamma$-ray band. The released high-energy electrons can diffuse through ISM and undergo inverse Compton scattering (ICS) with the background photon fields, making this a plausible scenario \citep{2020A&A...636A.113G,2022NatAs...6..199L}. For the interstellar radiation field of the ICS, we considered the CMB, near-infrared and far-infrared from interstellar dust and gas with standard value of $\rm T$ = 30 K and $u$ = 1 eV cm$^{-3}$ \citep{2006ApJ...648L..29P,2008ApJ...682..400P}. The target gas densities in Section \ref{sec:3} are adopted to calculate the contribution from bremsstrahlung in SrcA, SrcB and LHAASO source regions, respectively. 

\begin{figure}
    \centering
    \includegraphics[trim={0 0.cm 0 0}, clip, width=0.7\textwidth]{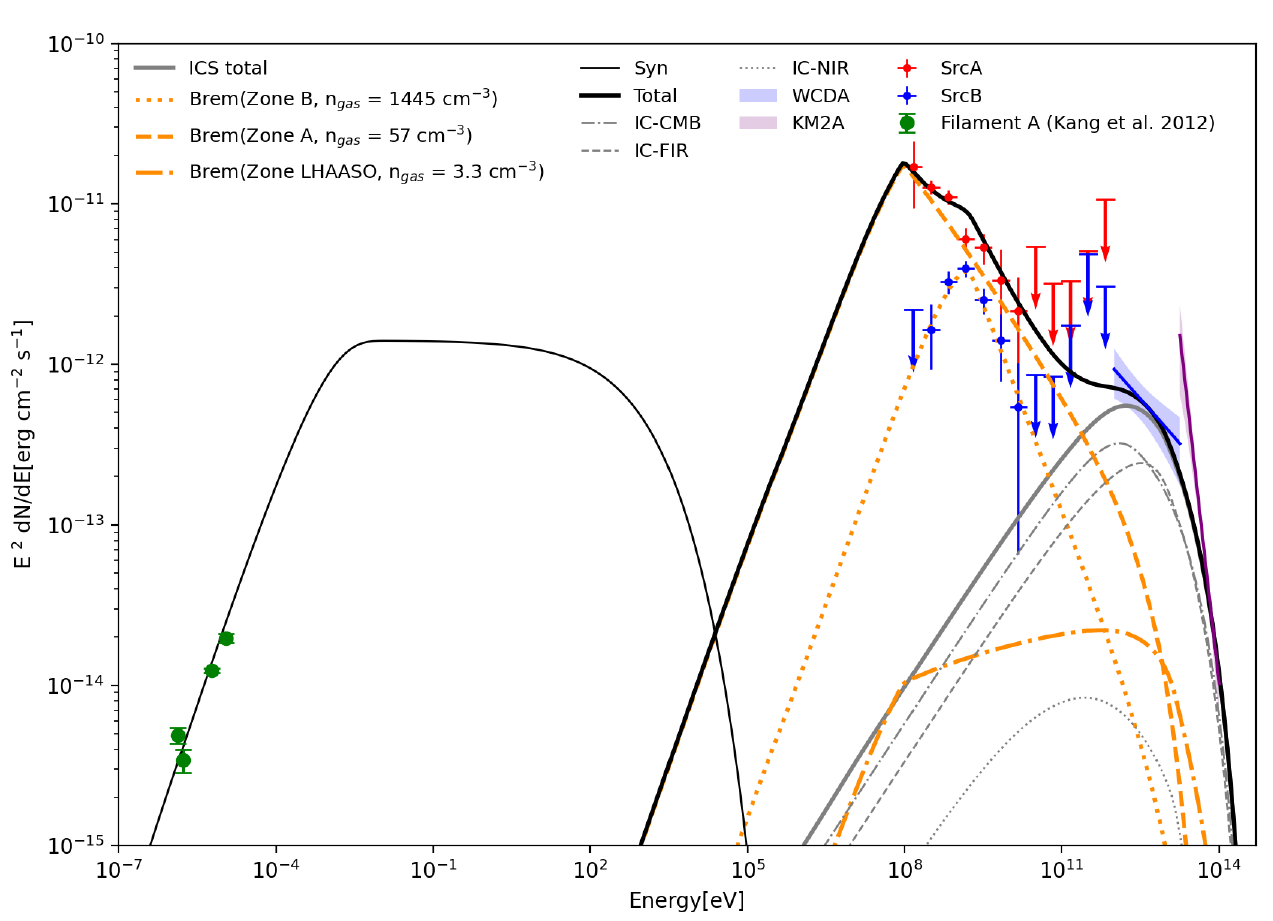}
    \caption{The leptonic model with bremsstrahlung-dominated GeV sources (different types of orange fit lines corresponding to different source regions, e.g., SrcA, SrcB and LHAASO source) and Inverse-Compton Scattering dominated LHAASO source (gray solid line). The blue and purple butterfly error region is subtracted from the LHAASO measurement \citep{2023arXiv230517030C}. The green data points in the radio energy band are calculated from \citet{2012AJ....143...75K}.}
    \label{fig:8}
\end{figure}

\begin{table*}[!htb]
	\centering
	\caption {Parameters for the leptonic model}
		{\begin{tabular}{cccccccc}
			\hline \hline
			\multirow{2}*{Zone} & \multirow{2}*{Model}  & $\alpha_{\rm e,1}$ &   $\alpha_{\rm e,2}$ &  E$_{\rm e, break}$  &  W$_{\rm e}$  &  E$_{\rm e, cut}$   & $\rm n_{gas}$  \\
            ~&~&~&~&~ & (erg) & (TeV) & (cm$^{-3}$)\\
			\hline
			\multirow{2}*{LHAASO} & \multirow{2}*{ICS}  &  \multirow{2}*{2.0} &  \multirow{2}*{3.0}  &  \multirow{2}*{$14.9^{+4.18}_{-3.72}$ TeV}  &  \multirow{2}*{$5.15(\rm d / 1.8 kpc)^{2} \times 10^{46}$} & \multirow{2}*{70.0}  & \multirow{2}*{3.3}   \\
            ~&Brem&~&~&~&~&~&~\\
			\hline
			SrcA & Brem &  1.46 &  $2.62^{+0.21}_{-0.13}$  &  $0.29^{+0.02}_{-0.05}$ GeV&   $7.79(\rm d / 1.8 kpc)^{2} \times 10^{47}$ & $10.8^{+2.9}_{-5.7} $  & 57  \\
			\hline
			SrcB & Brem  &  $-0.53^{+0.10}_{-0.06}$ &  $2.95^{+0.09}_{-0.07}$  &  $2.05^{+0.71}_{-0.45}$ GeV &  $2.07(\rm d / 1.8 kpc)^{2} \times 10^{45}$ & $21.3^{+8.4}_{-9.3}$  & 1445  \\
			\hline
			\hline 
	\end{tabular}}
	\tablecomments{$\alpha_{\rm e,1}$ and $\alpha_{\rm e,2}$ represent the indices below and above E$_{\rm e, break}$ for BPL spectrum of electrons. The total energy of electrons, $W_{\rm e}$, is calculated for $E_{\rm e} > 1$ GeV.}
	\label{tab:3}
\end{table*} 

In this scenario, the spectra of the electrons were assumed to be the broken power-law (ECBPL) with an exponential cutoff for typical PWNe, which follows \citep{2011MNRAS.410..381B,2019ApJ...874...50Z}:

\begin{equation}
\resizebox{0.6\hsize}{!}{$
\frac{dN_{\rm e}}{dE} \propto exp\left(-\frac{E}{E_{\rm e,cut}}\right) 
\begin{cases}
\left(\frac{E}{E_0}\right)^{-\alpha_{e1}} \qquad \qquad \qquad \qquad;  E < E_{\rm e,break}  \\
\left(\frac{E_{\rm e,break}}{E_0}\right)^{\alpha_{e2}-\alpha_{e1}} \left(\frac{E}{E_0}\right)^{-\alpha_{e2}} \,\,\,\,; E \geq E_{\rm e,break}
\end{cases}
$}
\end{equation}
\\
where $\alpha_{\rm e}$ and $\rm E_{e, cut}$ are the spectral index and the cutoff energy, $E_{\rm e,break}$ represents the break energy, combined with the radio flux density calculated from \citet{2012AJ....143...75K}.  Based on the gas analysis in Sect. \ref{sec:3}, we found that the very dense SrcB region would generate a high flux of bremsstrahlung component, which would be much higher than the ICS component if the same electronic distribution were considered. However, in contrast, the LHAASO source is located in a cavity of low-density gas distribution ($\sim$ 3.3 cm$^{-3}$), leading to the ICS process dominating the TeV flux contribution. Also, the Markov Chain Monte Carlo algorithm in the $naima$ package \citep{zabalza2015naima} to search the best-fit parameters for multi-wavelength data. It's worth noting that the radio data from \citet{2012AJ....143...75K} only extracted from part of SNR shell region, thus the synchrotron emission here would be the lowest prediction for the total SNR region, also, considering the radio spectra index $\sim$ 0.23 corresponding to $\sim$ 1.46 for electron spectrum, here we fixed 1.46 as the $\alpha_{\rm e,1}$ for SrcA. The fit results and parameters are summarized in Figure \ref{fig:8} and Table \ref{tab:3}.

In this model, SrcA and SrcB are assumed to originate from electrons accelerated by a supernova remnant (SNR), while the LHAASO TeV source might be associated with a potential pulsar halo. As a result, the spectral index of the LHAASO source differs from those of SrcA and SrcB. However, taking into account the fitting uncertainties, the spectral indices of SrcA and SrcB are relatively close to each other. This scenario also naturally explains why the LHAASO source appears point-like: the high-energy electrons produced in the LHAASO source cool rapidly, otherwise, the source would appear more extended. On the other hand, if the electrons responsible for the GeV emission originated from the LHAASO source, they would undergo inverse Compton scattering with background photon fields, producing a significant ICS contribution. This would result in a much harder GeV spectrum, which contradicts the observed soft spectrum discussed in Section \ref{sec:2.2}.

For LHAASO source region, we set $\alpha_{e1}$ = 2.0, and $\alpha_{e2}$ = $\alpha_{e1}$ + 1.0, and the synchrotron cooling timescale is equal to the age of SNR G172 (0.33 Myr) to reduce the number of free parameters (cooling time would be a function of energy, energy break and magnetic field strength would be related to each other), by fix the cut off energy E$_{\rm e, cut}$ = 70 TeV to explain the discontinuity between WCDA and KM2A spectra, further considering the E$_{\rm e, cut}$ = 1.25 $\times$ 10$^{\rm 7}$ t$_{\rm age; yr}^{-1}$ B$_{\rm \mu G}^{\rm -2}$ TeV, the magnetic field strength is constrained to a very low value of $\sim$ $1\ \mu$G, the best-fit E$_{\rm e, break}$ and others parameters are also summarized in Table \ref{tab:3}. For the SrcA and SrcB region, the $\gamma$-ray emission is dominated by the bremsstrahlung with different gas density, however, due to the absence of X-ray measurement results, the magnetic field strength is difficult to constrain, which makes it difficult to further constrain the electrons' cooling time and ICS contribution from these two sources. Furthermore, the GeV spectra from SrcB are much softer than the typical known $\gamma$-ray PWNe, for example Vela-X \citep{2013ApJ...774..110G} and HESS J1303-631 \citep{2012A&A...548A..46H}. Another possibility is that the LHAASO source is associated with a pulsar halo that has escaped from its host SNR \citep{2020A&A...636A.113G}, which could naturally explain the positional offset between SrcB and the LHAASO source. Furthermore, if we assume that SrcB is physically connected to the LHAASO source, and considering the presence of substantial gas along the line of sight between them, one would expect to observe GeV $\gamma$-ray emission along that path. However, as shown in Figure \ref{fig:1}, no such GeV emission structure is clearly detected. This discrepancy suggests that particle diffusion in this region may not be isotropic. A plausible explanation is the presence of filamentary structures that channel the propagation of high-energy particles, resulting in the observed $\gamma$-ray morphology in this work. This scenario is consistent with recent numerical simulations, which suggest that gamma-ray emission from pulsar halos can exhibit filamentary structures under certain conditions \citep{2018MNRAS.479.4526L}. Such structures make a physical connection between SrcB and the LHAASO source plausible. Future X-ray observations are required to constrain the X-ray flux in this region, and more detailed TeV observations would also be valuable.

\section{Conclusions}\label{sec:5}

We analyzed the GeV $\gamma$-ray emission in the vicinity of SNR candidate G172.8+1.5 using 16 yr of Fermi-LAT data. Two extended $\gamma$-ray sources were detected, named as SrcA and SrcB, respectively. Among them, SrcB has much softer photon spectra and spatial coincidence with a high-density molecular cloud, several OB stars and a star-forming region. SrcA is much more extended and spatially coincident with the SNR shell. Both hadronic and leptonic scenarios are discussed in this work. In the escaped CRs scenario, the $\gamma$-ray emission from SrcB can be interpreted as escaped CRs illuminating the surrounding MC. Bohm diffusion provides better fit performance compared with Kraichnan turbulence. However, in the much more extended SrcA region, the measured diffusion coefficient is significantly higher than that in the SrcB region, which is suggested to be dominated by a combination of Bohm diffusion near the SNR shell and Kraichnan turbulence in the interstellar medium region. In this scenario, TeV source 1LHAASO J0534+3533 is explained by the contribution from trapped ions, the extension discrepancy between GeV and TeV energy bands can be well explained with the model presented by \citet{2015MNRAS.447.2224B}. In YSCs scenario, the calculated smaller diffusion coefficient around SrcB is interpreted by the Bohm diffusion around the source region, which might be caused by CR-driven instabilities that amplify magnetic turbulence and suppress particle transport. In the leptonic scenario, the GeV emission observed in the high-density regions of SrcA and SrcB can be attributed to the bremsstrahlung process, with high-energy electrons accelerated by the associated SNR. The low-density environment of the LHAASO source region is dominated by inverse Compton scattering, and the high-energy electrons are suggested to be injected by a potential pulsar halo, their rapid cooling results in the point-like appearance of the LHAASO J0534+3533. Further details analysis from high-sensitivity detectors towards this region (such as CTA \citep{2013APh....43....3A}) is warranted.

\begin{acknowledgements}
We would like to thank Xiaona Sun, Houdun Zeng, P.P. Delia and Yang Su for invaluable discussions. This work is supported by the National Natural Science Foundation of China under the grants No. 12393853, 12375103, U1931204, 12103040, 12147208, and 12350610239.
\end{acknowledgements}

\bibliography{ref}

\begin{thebibliography}{}
\expandafter\ifx\csname natexlab\endcsname\relax\def\natexlab#1{#1}\fi
\providecommand{\url}[1]{\href{#1}{#1}}
\providecommand{\dodoi}[1]{doi:~\href{http://doi.org/#1}{\nolinkurl{#1}}}
\providecommand{\doeprint}[1]{\href{http://ascl.net/#1}{\nolinkurl{http://ascl.net/#1}}}
\providecommand{\doarXiv}[1]{\href{https://arxiv.org/abs/#1}{\nolinkurl{https://arxiv.org/abs/#1}}}

\bibitem[{{Abbott}(1982)}]{1982ApJ...259..282A}
{Abbott}, D.~C. 1982, \apj, 259, 282, \dodoi{10.1086/160166}

\bibitem[{{Abdollahi} {et~al.}(2020{\natexlab{a}}){Abdollahi}, {Acero},
  {Ackermann}, {Ajello}, {Atwood}, {Axelsson}, {Baldini}, {Ballet},
  {Barbiellini}, {Bastieri}, {Becerra Gonzalez}, {Bellazzini}, {Berretta},
  {Bissaldi}, {Blandford}, {Bloom}, {Bonino}, {Bottacini}, {Brandt}, {Bregeon},
  {Bruel}, {Buehler}, {Burnett}, {Buson}, {Cameron}, {Caputo}, {Caraveo},
  {Casandjian}, {Castro}, {Cavazzuti}, {Charles}, {Chaty}, {Chen}, {Cheung},
  {Chiaro}, {Ciprini}, {Cohen-Tanugi}, {Cominsky}, {Coronado-Bl{\'a}zquez},
  {Costantin}, {Cuoco}, {Cutini}, {D'Ammando}, {DeKlotz}, {de la Torre Luque},
  {de Palma}, {Desai}, {Digel}, {Di Lalla}, {Di Mauro}, {Di Venere},
  {Dom{\'\i}nguez}, {Dumora}, {Fana Dirirsa}, {Fegan}, {Ferrara},
  {Franckowiak}, {Fukazawa}, {Funk}, {Fusco}, {Gargano}, {Gasparrini},
  {Giglietto}, {Giommi}, {Giordano}, {Giroletti}, {Glanzman}, {Green},
  {Grenier}, {Griffin}, {Grondin}, {Grove}, {Guiriec}, {Harding}, {Hayashi},
  {Hays}, {Hewitt}, {Horan}, {J{\'o}hannesson}, {Johnson}, {Kamae}, {Kerr},
  {Kocevski}, {Kovac'evic'}, {Kuss}, {Landriu}, {Larsson}, {Latronico},
  {Lemoine-Goumard}, {Li}, {Liodakis}, {Longo}, {Loparco}, {Lott},
  {Lovellette}, {Lubrano}, {Madejski}, {Maldera}, {Malyshev}, {Manfreda},
  {Marchesini}, {Marcotulli}, {Mart{\'\i}-Devesa}, {Martin}, {Massaro},
  {Mazziotta}, {McEnery}, {Mereu}, {Meyer}, {Michelson}, {Mirabal}, {Mizuno},
  {Monzani}, {Morselli}, {Moskalenko}, {Negro}, {Nuss}, {Ojha}, {Omodei},
  {Orienti}, {Orlando}, {Ormes}, {Palatiello}, {Paliya}, {Paneque}, {Pei},
  {Pe{\~n}a-Herazo}, {Perkins}, {Persic}, {Pesce-Rollins}, {Petrosian},
  {Petrov}, {Piron}, {Poon}, {Porter}, {Principe}, {Rain{\`o}}, {Rando},
  {Razzano}, {Razzaque}, {Reimer}, {Reimer}, {Remy}, {Reposeur}, {Romani}, {Saz
  Parkinson}, {Schinzel}, {Serini}, {Sgr{\`o}}, {Siskind}, {Smith}, {Spandre},
  {Spinelli}, {Strong}, {Suson}, {Tajima}, {Takahashi}, {Tak}, {Thayer},
  {Thompson}, {Tibaldo}, {Torres}, {Torresi}, {Valverde}, {Van Klaveren}, {van
  Zyl}, {Wood}, {Yassine}, \& {Zaharijas}}]{Abdollahi2020a}
{Abdollahi}, S., {Acero}, F., {Ackermann}, M., {et~al.} 2020{\natexlab{a}},
  \apjs, 247, 33, \dodoi{10.3847/1538-4365/ab6bcb}

\bibitem[{{Abdollahi} {et~al.}(2020{\natexlab{b}}){Abdollahi}, {Acero},
  {Ackermann}, {Ajello}, {Atwood}, {Axelsson}, {Baldini}, {Ballet},
  {Barbiellini}, {Bastieri}, {Becerra Gonzalez}, {Bellazzini}, {Berretta},
  {Bissaldi}, {Blandford}, {Bloom}, {Bonino}, {Bottacini}, {Brandt}, {Bregeon},
  {Bruel}, {Buehler}, {Burnett}, {Buson}, {Cameron}, {Caputo}, {Caraveo},
  {Casandjian}, {Castro}, {Cavazzuti}, {Charles}, {Chaty}, {Chen}, {Cheung},
  {Chiaro}, {Ciprini}, {Cohen-Tanugi}, {Cominsky}, {Coronado-Bl{\'a}zquez},
  {Costantin}, {Cuoco}, {Cutini}, {D'Ammando}, {DeKlotz}, {de la Torre Luque},
  {de Palma}, {Desai}, {Digel}, {Di Lalla}, {Di Mauro}, {Di Venere},
  {Dom{\'\i}nguez}, {Dumora}, {Fana Dirirsa}, {Fegan}, {Ferrara},
  {Franckowiak}, {Fukazawa}, {Funk}, {Fusco}, {Gargano}, {Gasparrini},
  {Giglietto}, {Giommi}, {Giordano}, {Giroletti}, {Glanzman}, {Green},
  {Grenier}, {Griffin}, {Grondin}, {Grove}, {Guiriec}, {Harding}, {Hayashi},
  {Hays}, {Hewitt}, {Horan}, {J{\'o}hannesson}, {Johnson}, {Kamae}, {Kerr},
  {Kocevski}, {Kovac'evic'}, {Kuss}, {Landriu}, {Larsson}, {Latronico},
  {Lemoine-Goumard}, {Li}, {Liodakis}, {Longo}, {Loparco}, {Lott},
  {Lovellette}, {Lubrano}, {Madejski}, {Maldera}, {Malyshev}, {Manfreda},
  {Marchesini}, {Marcotulli}, {Mart{\'\i}-Devesa}, {Martin}, {Massaro},
  {Mazziotta}, {McEnery}, {Mereu}, {Meyer}, {Michelson}, {Mirabal}, {Mizuno},
  {Monzani}, {Morselli}, {Moskalenko}, {Negro}, {Nuss}, {Ojha}, {Omodei},
  {Orienti}, {Orlando}, {Ormes}, {Palatiello}, {Paliya}, {Paneque}, {Pei},
  {Pe{\~n}a-Herazo}, {Perkins}, {Persic}, {Pesce-Rollins}, {Petrosian},
  {Petrov}, {Piron}, {Poon}, {Porter}, {Principe}, {Rain{\`o}}, {Rando},
  {Razzano}, {Razzaque}, {Reimer}, {Reimer}, {Remy}, {Reposeur}, {Romani}, {Saz
  Parkinson}, {Schinzel}, {Serini}, {Sgr{\`o}}, {Siskind}, {Smith}, {Spandre},
  {Spinelli}, {Strong}, {Suson}, {Tajima}, {Takahashi}, {Tak}, {Thayer},
  {Thompson}, {Tibaldo}, {Torres}, {Torresi}, {Valverde}, {Van Klaveren}, {van
  Zyl}, {Wood}, {Yassine}, \& {Zaharijas}}]{2020ApJS..247...33A}
---. 2020{\natexlab{b}}, \apjs, 247, 33, \dodoi{10.3847/1538-4365/ab6bcb}

\bibitem[{{Abeysekara} {et~al.}(2021){Abeysekara}, {Albert}, {Alfaro},
  {Alvarez}, {Camacho}, {Arteaga-Vel{\'a}zquez}, {Arunbabu}, {Rojas},
  {Solares}, {Baghmanyan}, {Belmont-Moreno}, {BenZvi}, {Blandford}, {Brisbois},
  {Caballero-Mora}, {Capistr{\'a}n}, {Carrami{\~n}ana}, {Casanova}, {Cotti},
  {Le{\'o}n}, {De la Fuente}, {Hernandez}, {Dingus}, {DuVernois}, {Durocher},
  {D{\'\i}az-V{\'e}lez}, {Ellsworth}, {Engel}, {Espinoza}, {Fan}, {Fang},
  {Fleischhack}, {Fraija}, {Galv{\'a}n-G{\'a}mez}, {Garcia},
  {Garc{\'\i}a-Gonz{\'a}lez}, {Garfias}, {Giacinti}, {Gonz{\'a}lez}, {Goodman},
  {Harding}, {Hernandez}, {Hinton}, {Hona}, {Huang}, {Hueyotl-Zahuantitla},
  {H{\"u}ntemeyer}, {Iriarte}, {Jardin-Blicq}, {Joshi}, {Kieda}, {Lara}, {Lee},
  {Vargas}, {Linnemann}, {Longinotti}, {Luis-Raya}, {Lundeen}, {Malone},
  {Martinez}, {Martinez-Castellanos}, {Mart{\'\i}nez-Castro}, {Matthews},
  {Miranda-Romagnoli}, {Morales-Soto}, {Moreno}, {Mostaf{\'a}}, {Nayerhoda},
  {Nellen}, {Newbold}, {Nisa}, {Noriega-Papaqui}, {Olivera-Nieto}, {Omodei},
  {Peisker}, {P{\'e}rez Araujo}, {P{\'e}rez-P{\'e}rez}, {Ren}, {Rho},
  {Rosa-Gonz{\'a}lez}, {Ruiz-Velasco}, {Salazar}, {Greus}, {Sandoval},
  {Schneider}, {Schoorlemmer}, {Serna}, {Smith}, {Springer}, {Surajbali},
  {Tollefson}, {Torres}, {Torres-Escobedo}, {Ure{\~n}a-Mena}, {Weisgarber},
  {Werner}, {Willox}, {Zepeda}, {Zhou}, {De Le{\'o}n}, \&
  {{\'A}lvarez}}]{2021NatAs...5..465A}
{Abeysekara}, A.~U., {Albert}, A., {Alfaro}, R., {et~al.} 2021, Nature
  Astronomy, 5, 465, \dodoi{10.1038/s41550-021-01318-y}

\bibitem[{{Abramowski} {et~al.}(2012){Abramowski}, {Acero}, {Aharonian},
  {Akhperjanian}, {Anton}, {Balzer}, {Barnacka}, {Barres de Almeida},
  {Becherini}, {Becker}, {Behera}, {Bernl{\"o}hr}, {Birsin}, {Biteau},
  {Bochow}, {Boisson}, {Bolmont}, {Bordas}, {Brucker}, {Brun}, {Brun}, {Bulik},
  {B{\"u}sching}, {Carrigan}, {Casanova}, {Cerruti}, {Chadwick}, {Charbonnier},
  {Chaves}, {Cheesebrough}, {Chounet}, {Clapson}, {Coignet}, {Cologna},
  {Conrad}, {Dalton}, {Daniel}, {Davids}, {Degrange}, {Deil}, {Dickinson},
  {Djannati-Ata{\"\i}}, {Domainko}, {O'C. Drury}, {Dubois}, {Dubus}, {Dutson},
  {Dyks}, {Dyrda}, {Egberts}, {Eger}, {Espigat}, {Fallon}, {Farnier}, {Fegan},
  {Feinstein}, {Fernandes}, {Fiasson}, {Fontaine}, {F{\"o}rster},
  {F{\"u}{\ss}ling}, {Gallant}, {Gast}, {G{\'e}rard}, {Gerbig}, {Giebels},
  {Glicenstein}, {Gl{\"u}ck}, {Goret}, {G{\"o}ring}, {H{\"a}ffner}, {Hague},
  {Hampf}, {Hauser}, {Heinz}, {Heinzelmann}, {Henri}, {Hermann}, {Hinton},
  {Hoffmann}, {Hofmann}, {Hofverberg}, {Holler}, {Horns}, {Jacholkowska}, {de
  Jager}, {Jahn}, {Jamrozy}, {Jung}, {Kastendieck}, {Katarzy{\'n}ski}, {Katz},
  {Kaufmann}, {Keogh}, {Khangulyan}, {Kh{\'e}lifi}, {Klochkov}, {Klu{\.z}niak},
  {Kneiske}, {Komin}, {Kosack}, {Kossakowski}, {Laffon}, {Lamanna}, {Lennarz},
  {Lohse}, {Lopatin}, {Lu}, {Marandon}, {Marcowith}, {Masbou}, {Maurin},
  {Maxted}, {Mayer}, {McComb}, {Medina}, {M{\'e}hault}, {Moderski}, {Moulin},
  {Naumann}, {Naumann-Godo}, {de Naurois}, {Nedbal}, {Nekrassov}, {Nguyen},
  {Nicholas}, {Niemiec}, {Nolan}, {Ohm}, {de O{\~n}a Wilhelmi}, {Opitz},
  {Ostrowski}, {Oya}, {Panter}, {Paz Arribas}, {Pedaletti}, {Pelletier},
  {Petrucci}, {Pita}, {P{\"u}hlhofer}, {Punch}, {Quirrenbach}, {Raue},
  {Rayner}, {Reimer}, {Reimer}, {Renaud}, {de Los Reyes}, {Rieger}, {Ripken},
  {Rob}, {Rosier-Lees}, {Rowell}, {Rudak}, {Rulten}, {Ruppel}, {Sahakian},
  {Sanchez}, {Santangelo}, {Schlickeiser}, {Sch{\"o}ck}, {Schulz}, {Schwanke},
  {Schwarzburg}, {Schwemmer}, {Sheidaei}, {Sikora}, {Skilton}, {Sol},
  {Spengler}, {Stawarz}, {Steenkamp}, {Stegmann}, {Stinzing}, {Stycz},
  {Sushch}, {Szostek}, {Tavernet}, {Terrier}, {Tluczykont}, {Valerius}, {van
  Eldik}, {Vasileiadis}, {Venter}, {Vialle}, {Viana}, {Vincent}, {V{\"o}lk},
  {Volpe}, {Vorobiov}, {Vorster}, {Wagner}, {Ward}, {White}, {Wierzcholska},
  {Zacharias}, {Zajczyk}, {Zdziarski}, {Zech}, \&
  {Zechlin}}]{2012A&A...537A.114A}
{Abramowski}, A., {Acero}, F., {Aharonian}, F., {et~al.} 2012, \aap, 537, A114,
  \dodoi{10.1051/0004-6361/201117928}

\bibitem[{{Acero} {et~al.}(2016){Acero}, {Ackermann}, {Ajello}, {Albert},
  {Baldini}, {Ballet}, {Barbiellini}, {Bastieri}, {Bellazzini}, {Bissaldi},
  {Bloom}, {Bonino}, {Bottacini}, {Brandt}, {Bregeon}, {Bruel}, {Buehler},
  {Buson}, {Caliandro}, {Cameron}, {Caragiulo}, {Caraveo}, {Casandjian},
  {Cavazzuti}, {Cecchi}, {Charles}, {Chekhtman}, {Chiang}, {Chiaro}, {Ciprini},
  {Claus}, {Cohen-Tanugi}, {Conrad}, {Cuoco}, {Cutini}, {D'Ammando}, {de
  Angelis}, {de Palma}, {Desiante}, {Digel}, {Di Venere}, {Drell}, {Favuzzi},
  {Fegan}, {Ferrara}, {Focke}, {Franckowiak}, {Funk}, {Fusco}, {Gargano},
  {Gasparrini}, {Giglietto}, {Giordano}, {Giroletti}, {Glanzman}, {Godfrey},
  {Grenier}, {Guiriec}, {Hadasch}, {Harding}, {Hayashi}, {Hays}, {Hewitt},
  {Hill}, {Horan}, {Hou}, {Jogler}, {J{\'o}hannesson}, {Kamae}, {Kuss},
  {Landriu}, {Larsson}, {Latronico}, {Li}, {Li}, {Longo}, {Loparco},
  {Lovellette}, {Lubrano}, {Maldera}, {Malyshev}, {Manfreda}, {Martin},
  {Mayer}, {Mazziotta}, {McEnery}, {Michelson}, {Mirabal}, {Mizuno}, {Monzani},
  {Morselli}, {Nuss}, {Ohsugi}, {Omodei}, {Orienti}, {Orlando}, {Ormes},
  {Paneque}, {Pesce-Rollins}, {Piron}, {Pivato}, {Rain{\`o}}, {Rando},
  {Razzano}, {Razzaque}, {Reimer}, {Reimer}, {Remy}, {Renault},
  {S{\'a}nchez-Conde}, {Schaal}, {Schulz}, {Sgr{\`o}}, {Siskind}, {Spada},
  {Spandre}, {Spinelli}, {Strong}, {Suson}, {Tajima}, {Takahashi}, {Thayer},
  {Thompson}, {Tibaldo}, {Tinivella}, {Torres}, {Tosti}, {Troja}, {Vianello},
  {Werner}, {Wood}, {Wood}, {Zaharijas}, \& {Zimmer}}]{2016ApJS..223...26A}
{Acero}, F., {Ackermann}, M., {Ajello}, M., {et~al.} 2016, \apjs, 223, 26,
  \dodoi{10.3847/0067-0049/223/2/26}

\bibitem[{{Acharya} {et~al.}(2013){Acharya}, {Actis}, {Aghajani}, {Agnetta},
  {Aguilar}, {Aharonian}, {Ajello}, {Akhperjanian}, {Alcubierre},
  {Aleksi{\'c}}, {Alfaro}, {Aliu}, {Allafort}, {Allan}, {Allekotte}, {Amato},
  {Anderson}, {Ang{\"u}ner}, {Antonelli}, {Antoranz}, {Aravantinos}, {Arlen},
  {Armstrong}, {Arnaldi}, {Arrabito}, {Asano}, {Ashton}, {Asorey}, {Awane},
  {Baba}, {Babic}, {Baby}, {B{\"a}hr}, {Bais}, {Baixeras}, {Bajtlik}, {Balbo},
  {Balis}, {Balkowski}, {Bamba}, {Bandiera}, {Barber}, {Barbier},
  {Barcel{\'o}}, {Barnacka}, {Barnstedt}, {Barres de Almeida}, {Barrio},
  {Basili}, {Basso}, {Bastieri}, {Bauer}, {Baushev}, {Becerra}, {Becherini},
  {Bechtol}, {Becker Tjus}, {Beckmann}, {Bednarek}, {Behera}, {Belluso},
  {Benbow}, {Berdugo}, {Berger}, {Bernard}, {Bernardino}, {Bernl{\"o}hr},
  {Bhat}, {Bhattacharyya}, {Bigongiari}, {Biland}, {Billotta}, {Bird},
  {Birsin}, {Bissaldi}, {Biteau}, {Bitossi}, {Blake}, {Blanch Bigas}, {Blasi},
  {Bobkov}, {Boccone}, {Boettcher}, {Bogacz}, {Bogart}, {Bogdan}, {Boisson},
  {Boix Gargallo}, {Bolmont}, {Bonanno}, {Bonardi}, {Bonev}, {Bonifacio},
  {Bonnoli}, {Bordas}, {Borgland}, {Borkowski}, {Bose}, {Botner}, {Bottani},
  {Bouchet}, {Bourgeat}, {Boutonnet}, {Bouvier}, {Brau-Nogu{\'e}}, {Braun},
  {Bretz}, {Briggs}, {Bringmann}, {Brook}, {Brun}, {Brunetti}, {Buanes},
  {Buckley}, {Buehler}, {Bugaev}, {Bulgarelli}, {Bulik}, {Busetto}, {Buson},
  {Byrum}, {Cailles}, {Cameron}, {Camprecios}, {Canestrari}, {Cantu},
  {Capalbi}, {Caraveo}, {Carmona}, {Carosi}, {Carr}, {Carton}, {Casanova},
  {Casiraghi}, {Catalano}, {Cavazzani}, {Cazaux}, {Cerruti}, {Chabanne},
  {Chadwick}, {Champion}, {Chen}, {Chiang}, {Chiappetti}, {Chikawa}, {Chitnis},
  {Chollet}, {Chudoba}, {Cie{\'s}lar}, {Cillis}, {Cohen-Tanugi},
  {Colafrancesco}, {Colin}, {Colome}, {Colonges}, {Compin}, {Conconi},
  {Conforti}, {Connaughton}, {Conrad}, {Contreras}, {Coppi}, {Corona}, {Corti},
  {Cortina}, {Cossio}, {Costantini}, {Cotter}, {Courty}, {Couturier}, {Covino},
  {Crimi}, {Criswell}, {Croston}, {Cusumano}, {Dafonseca}, {Dale}, {Daniel},
  {Darling}, {Davids}, {Dazzi}, {De Angelis}, {De Caprio}, {De Frondat}, {de
  Gouveia Dal Pino}, {de la Calle}, {De La Vega}, {de los Reyes Lopez}, {De
  Lotto}, {De Luca}, {de Mello Neto}, {de Naurois}, {de Oliveira}, {de O{\~n}a
  Wilhelmi}, {de Souza}, {Decerprit}, {Decock}, {Deil}, {Delagnes}, \&
  {Deleglise}}]{2013APh....43....3A}
{Acharya}, B.~S., {Actis}, M., {Aghajani}, T., {et~al.} 2013, Astroparticle
  Physics, 43, 3, \dodoi{10.1016/j.astropartphys.2013.01.007}

\bibitem[{{Ackermann} {et~al.}(2011){Ackermann}, {Ajello}, {Allafort},
  {Baldini}, {Ballet}, {Barbiellini}, {Bastieri}, {Belfiore}, {Bellazzini},
  {Berenji}, {Blandford}, {Bloom}, {Bonamente}, {Borgland}, {Bottacini},
  {Brigida}, {Bruel}, {Buehler}, {Buson}, {Caliandro}, {Cameron}, {Caraveo},
  {Casandjian}, {Cecchi}, {Chekhtman}, {Cheung}, {Chiang}, {Ciprini}, {Claus},
  {Cohen-Tanugi}, {de Angelis}, {de Palma}, {Dermer}, {do Couto e Silva},
  {Drell}, {Dumora}, {Favuzzi}, {Fegan}, {Focke}, {Fortin}, {Fukazawa},
  {Fusco}, {Gargano}, {Germani}, {Giglietto}, {Giordano}, {Giroletti},
  {Glanzman}, {Godfrey}, {Grenier}, {Guillemot}, {Guiriec}, {Hadasch},
  {Hanabata}, {Harding}, {Hayashida}, {Hayashi}, {Hays}, {J{\'o}hannesson},
  {Johnson}, {Kamae}, {Katagiri}, {Kataoka}, {Kerr}, {Kn{\"o}dlseder}, {Kuss},
  {Lande}, {Latronico}, {Lee}, {Longo}, {Loparco}, {Lott}, {Lovellette},
  {Lubrano}, {Martin}, {Mazziotta}, {McEnery}, {Mehault}, {Michelson},
  {Mitthumsiri}, {Mizuno}, {Monte}, {Monzani}, {Morselli}, {Moskalenko},
  {Murgia}, {Naumann-Godo}, {Nolan}, {Norris}, {Nuss}, {Ohsugi}, {Okumura},
  {Orlando}, {Ormes}, {Ozaki}, {Paneque}, {Parent}, {Pesce-Rollins},
  {Pierbattista}, {Piron}, {Pohl}, {Prokhorov}, {Rain{\`o}}, {Rando},
  {Razzano}, {Reposeur}, {Ritz}, {Parkinson}, {Sgr{\`o}}, {Siskind}, {Smith},
  {Spinelli}, {Strong}, {Takahashi}, {Tanaka}, {Thayer}, {Thayer}, {Thompson},
  {Tibaldo}, {Torres}, {Tosti}, {Tramacere}, {Troja}, {Uchiyama},
  {Vandenbroucke}, {Vasileiou}, {Vianello}, {Vitale}, {Waite}, {Wang}, {Winer},
  {Wood}, {Yang}, {Zimmer}, \& {Bontemps}}]{2011Sci...334.1103A}
{Ackermann}, M., {Ajello}, M., {Allafort}, A., {et~al.} 2011, Science, 334,
  1103, \dodoi{10.1126/science.1210311}

\bibitem[{{Aguilar} {et~al.}(2015){Aguilar}, {Aisa}, {Alpat}, {Alvino},
  {Ambrosi}, {Andeen}, {Arruda}, {Attig}, {Azzarello}, {Bachlechner}, {Barao},
  {Barrau}, {Barrin}, {Bartoloni}, {Basara}, {Battarbee}, {Battiston}, {Bazo},
  {Becker}, {Behlmann}, {Beischer}, {Berdugo}, {Bertucci}, {Bindi},
  {Bizzaglia}, {Bizzarri}, {Boella}, {de Boer}, {Bollweg}, {Bonnivard},
  {Borgia}, {Borsini}, {Boschini}, {Bourquin}, {Burger}, {Cadoux}, {Cai},
  {Capell}, {Caroff}, {Casaus}, {Castellini}, {Cernuda}, {Cerreta}, {Cervelli},
  {Chae}, {Chang}, {Chen}, {Chen}, {Chen}, {Chen}, {Cheng}, {Chou},
  {Choumilov}, {Choutko}, {Chung}, {Clark}, {Clavero}, {Coignet}, {Consolandi},
  {Contin}, {Corti}, {Gil}, {Coste}, {Creus}, {Crispoltoni}, {Cui}, {Dai},
  {Delgado}, {Della Torre}, {Demirk{\"o}z}, {Derome}, {Di Falco}, {Di Masso},
  {Dimiccoli}, {D{\'\i}az}, {von Doetinchem}, {Donnini}, {Duranti}, {D'Urso},
  {Egorov}, {Eline}, {Eppling}, {Eronen}, {Fan}, {Farnesini}, {Feng},
  {Fiandrini}, {Fiasson}, {Finch}, {Fisher}, {Formato}, {Galaktionov},
  {Gallucci}, {Garc{\'\i}a}, {Garc{\'\i}a-L{\'o}pez}, {Gargiulo}, {Gast},
  {Gebauer}, {Gervasi}, {Ghelfi}, {Giovacchini}, {Goglov}, {Gong}, {Goy},
  {Grabski}, {Grandi}, {Graziani}, {Guandalini}, {Guerri}, {Guo}, {Haas},
  {Habiby}, {Haino}, {Han}, {He}, {Heil}, {Hoffman}, {Hsieh}, {Huang}, {Huh},
  {Incagli}, {Ionica}, {Jang}, {Jinchi}, {Kanishev}, {Kim}, {Kim}, {Kirn},
  {Korkmaz}, {Kossakowski}, {Kounina}, {Kounine}, {Koutsenko}, {Krafczyk}, {La
  Vacca}, {Laudi}, {Laurenti}, {Lazzizzera}, {Lebedev}, {Lee}, {Lee}, {Leluc},
  {Li}, {Li}, {Li}, {Li}, {Li}, {Li}, {Li}, {Li}, {Li}, {Li}, {Lim}, {Lin},
  {Lipari}, {Lippert}, {Liu}, {Liu}, {Liu}, {Lolli}, {Lomtadze}, {Lu}, {Lu},
  {Lu}, {Luebelsmeyer}, {Luo}, {Luo}, {Lv}, {Majka}, {Ma{\~n}{\'a}},
  {Mar{\'\i}n}, {Martin}, {Mart{\'\i}nez}, {Masi}, {Maurin}, {Menchaca-Rocha},
  {Meng}, {Mo}, {Morescalchi}, {Mott}, {M{\"u}ller}, {Nelson}, {Ni}, {Nikonov},
  {Nozzoli}, {Nunes}, {Obermeier}, {Oliva}, {Orcinha}, {Palmonari},
  {Palomares}, {Paniccia}, {Papi}, {Pauluzzi}, {Pedreschi}, {Pensotti},
  {Pereira}, {Picot-Clemente}, {Pilo}, \& {Piluso}}]{2015PhRvL.115u1101A}
{Aguilar}, M., {Aisa}, D., {Alpat}, B., {et~al.} 2015, \prl, 115, 211101,
  \dodoi{10.1103/PhysRevLett.115.211101}

\bibitem[{{Aharonian} {et~al.}(2019){Aharonian}, {Yang}, \& {de O{\~n}a
  Wilhelmi}}]{2019NatAs...3..561A}
{Aharonian}, F., {Yang}, R., \& {de O{\~n}a Wilhelmi}, E. 2019, Nature
  Astronomy, 3, 561, \dodoi{10.1038/s41550-019-0724-0}

\bibitem[{{Aharonian} {et~al.}(2022){Aharonian}, {Ashkar}, {Backes}, {Barbosa
  Martins}, {Becherini}, {Berge}, {Bi}, {B{\"o}ttcher}, {de Bony de Lavergne},
  {Bradascio}, {Brose}, {Brun}, {Bulik}, {Burger-Scheidlin}, {Cangemi},
  {Caroff}, {Casanova}, {Cerruti}, {Chand}, {Chandra}, {Chen}, {Chibueze},
  {Cristofari}, {Damascene Mbarubucyeye}, {Djannati-Ata{\"\i}}, {Ernenwein},
  {Feijen}, {Fichet de Clairfontaine}, {Fontaine}, {Funk}, {Gabici}, {Gallant},
  {Ghafourizadeh}, {Giavitto}, {Giunti}, {Glawion}, {Glicenstein}, {Goswami},
  {Grondin}, {H{\"a}rer}, {Haupt}, {Hinton}, {H{\"o}rbe}, {Hofmann}, {Holch},
  {Holler}, {Horns}, {Jamrozy}, {Joshi}, {Jung-Richardt}, {Kasai},
  {Katarzy{\'n}ski}, {Katz}, {Kh{\'e}lifi}, {Klu{\'z}niak}, {Komin}, {Kosack},
  {Kostunin}, {Kukec Mezek}, {Lang}, {Le Stum}, {Lemi{\`e}re},
  {Lemoine-Goumard}, {Lenain}, {Leuschner}, {Lohse}, {Luashvili}, {Lypova},
  {Mackey}, {Majumdar}, {Malyshev}, {Marandon}, {Marchegiani}, {Marcowith},
  {Mart{\'\i}-Devesa}, {Marx}, {Maurin}, {Meyer}, {Mitchell}, {Moderski},
  {Mohrmann}, {Montanari}, {Moulin}, {Muller}, {Murach}, {Nakashima}, {de
  Naurois}, {Nayerhoda}, {Niemiec}, {Ohm}, {Olivera-Nieto}, {de Ona Wilhelmi},
  {Ostrowski}, {Panny}, {Panter}, {Parsons}, {Peron}, {Prokhorov},
  {P{\"u}hlhofer}, {Punch}, {Quirrenbach}, {Rauth}, {Reichherzer}, {Reimer},
  {Reimer}, {Renaud}, {Reville}, {Rieger}, {Rowell}, {Rudak}, {Ruiz-Velasco},
  {Sahakian}, {Salzmann}, {Sanchez}, {Santangelo}, {Sasaki}, {Sch{\"u}ssler},
  {Schutte}, {Schwanke}, {Shapopi}, {Specovius}, {Spencer}, {Stawarz},
  {Steenkamp}, {Steinmassl}, {Steppa}, {Sushch}, {Suzuki}, {Takahashi},
  {Tanaka}, {Terrier}, {Thorpe-Morgan}, {Tsirou}, {Tsuji}, {Tuffs}, {Unbehaun},
  {van Eldik}, {van Soelen}, {Vecchi}, {Veh}, {Venter}, {Vink}, {Wagner},
  {White}, {Wierzcholska}, {Wong}, {Zacharias}, {Zargaryan}, {Zdziarski},
  {Zhu}, {Zouari}, {{\.Z}ywucka}, {Blackwell}, {Braiding}, {Burton}, {Cubuk},
  {Filipovi{\'c}}, {Tothill}, \& {Wong}}]{2022A&A...666A.124A}
{Aharonian}, F., {Ashkar}, H., {Backes}, M., {et~al.} 2022, \aap, 666, A124,
  \dodoi{10.1051/0004-6361/202244323}

\bibitem[{{Akaike}(1974)}]{1974AIC}
{Akaike}, H. 1974, IEEE Transactions on Automatic Control, 19, 716

\bibitem[{{Albert} {et~al.}(2020){Albert}, {Alfaro}, {Alvarez}, {Camacho},
  {Arteaga-Vel{\'a}zquez}, {Arunbabu}, {Avila Rojas}, {Ayala Solares},
  {Baghmanyan}, {Belmont-Moreno}, {BenZvi}, {Brisbois}, {Caballero-Mora},
  {Capistr{\'a}n}, {Carrami{\~n}ana}, {Casanova}, {Cotti}, {Couti{\~n}o de
  Le{\'o}n}, {De la Fuente}, {Diaz Hernandez}, {Diaz-Cruz}, {Dingus},
  {DuVernois}, {Durocher}, {D{\'\i}az-V{\'e}lez}, {Ellsworth}, {Engel},
  {Espinoza}, {Fan}, {Fang}, {Alonso}, {Fleischhack}, {Fraija},
  {Galv{\'a}n-G{\'a}mez}, {Garcia}, {Garc{\'\i}a-Gonz{\'a}lez}, {Garfias},
  {Giacinti}, {Gonz{\'a}lez}, {Goodman}, {Harding}, {Hernandez}, {Hinton},
  {Hona}, {Huang}, {Hueyotl-Zahuantitla}, {H{\"u}ntemeyer}, {Iriarte},
  {Jardin-Blicq}, {Joshi}, {Kieda}, {Lara}, {Lee}, {Le{\'o}n Vargas},
  {Linnemann}, {Longinotti}, {Luis-Raya}, {Lundeen}, {L{\'o}pez-Coto},
  {Malone}, {Marandon}, {Martinez}, {Martinez-Castellanos},
  {Mart{\'\i}nez-Castro}, {Matthews}, {Miranda-Romagnoli}, {Morales-Soto},
  {Moreno}, {Mostaf{\'a}}, {Nayerhoda}, {Nellen}, {Newbold}, {Nisa},
  {Noriega-Papaqui}, {Olivera-Nieto}, {Omodei}, {Peisker}, {P{\'e}rez Araujo},
  {P{\'e}rez-P{\'e}rez}, {Ren}, {Rho}, {Rivi{\`e}re}, {Rosa-Gonz{\'a}lez},
  {Ruiz-Velasco}, {Salazar}, {Salesa Greus}, {Sandoval}, {Schneider},
  {Schoorlemmer}, {Serna}, {Sinnis}, {Smith}, {Springer}, {Surajbali},
  {Tollefson}, {Torres}, {Torres-Escobedo}, {Ukwatta}, {Ure{\~n}a-Mena},
  {Weisgarber}, {Werner}, {Willox}, {Zepeda}, {Zhou}, {de Le{\'o}n},
  {{\'A}lvarez}, \& {HAWC Collaboration}}]{2020ApJ...905...76A}
{Albert}, A., {Alfaro}, R., {Alvarez}, C., {et~al.} 2020, \apj, 905, 76,
  \dodoi{10.3847/1538-4357/abc2d8}

\bibitem[{{Anderson} {et~al.}(2014){Anderson}, {Bania}, {Balser}, {Cunningham},
  {Wenger}, {Johnstone}, \& {Armentrout}}]{2014ApJS..212....1A}
{Anderson}, L.~D., {Bania}, T.~M., {Balser}, D.~S., {et~al.} 2014, \apjs, 212,
  1, \dodoi{10.1088/0067-0049/212/1/1}

\bibitem[{{Ballet} {et~al.}(2023){Ballet}, {Bruel}, {Burnett}, {Lott}, \& {The
  Fermi-LAT collaboration}}]{2023arXiv230712546B}
{Ballet}, J., {Bruel}, P., {Burnett}, T.~H., {Lott}, B., \& {The Fermi-LAT
  collaboration}. 2023, arXiv e-prints, arXiv:2307.12546,
  \dodoi{10.48550/arXiv.2307.12546}

\bibitem[{{Bell}(2004)}]{2004MNRAS.353..550B}
{Bell}, A.~R. 2004, \mnras, 353, 550, \dodoi{10.1111/j.1365-2966.2004.08097.x}

\bibitem[{{Bell}(2015)}]{2015MNRAS.447.2224B}
---. 2015, \mnras, 447, 2224, \dodoi{10.1093/mnras/stu2596}

\bibitem[{{Bica} {et~al.}(2003){Bica}, {Dutra}, \&
  {Barbuy}}]{2003A&A...397..177B}
{Bica}, E., {Dutra}, C.~M., \& {Barbuy}, B. 2003, \aap, 397, 177,
  \dodoi{10.1051/0004-6361:20021479}

\bibitem[{{Blasi}(2013)}]{2013A&ARv..21...70B}
{Blasi}, P. 2013, \aapr, 21, 70, \dodoi{10.1007/s00159-013-0070-7}

\bibitem[{{Blitz} {et~al.}(1982){Blitz}, {Fich}, \&
  {Stark}}]{1982ApJS...49..183B}
{Blitz}, L., {Fich}, M., \& {Stark}, A.~A. 1982, \apjs, 49, 183,
  \dodoi{10.1086/190795}

\bibitem[{Bohm(1949)}]{1949bohm}
Bohm, D. 1949, (McGraw-Hill, New York)

\bibitem[{{Bolatto} {et~al.}(2013){Bolatto}, {Wolfire}, \&
  {Leroy}}]{bolatto2013}
{Bolatto}, A.~D., {Wolfire}, M., \& {Leroy}, A.~K. 2013, \araa, 51, 207,
  \dodoi{10.1146/annurev-astro-082812-140944}

\bibitem[{{Bucciantini} {et~al.}(2011){Bucciantini}, {Arons}, \&
  {Amato}}]{2011MNRAS.410..381B}
{Bucciantini}, N., {Arons}, J., \& {Amato}, E. 2011, \mnras, 410, 381,
  \dodoi{10.1111/j.1365-2966.2010.17449.x}

\bibitem[{{Buckner} \& {Froebrich}(2013)}]{2013MNRAS.436.1465B}
{Buckner}, A. S.~M., \& {Froebrich}, D. 2013, \mnras, 436, 1465,
  \dodoi{10.1093/mnras/stt1665}

\bibitem[{{Bykov} {et~al.}(2020){Bykov}, {Marcowith}, {Amato}, {Kalyashova},
  {Kruijssen}, \& {Waxman}}]{2020SSRv..216...42B}
{Bykov}, A.~M., {Marcowith}, A., {Amato}, E., {et~al.} 2020, \ssr, 216, 42,
  \dodoi{10.1007/s11214-020-00663-0}

\bibitem[{{Camargo} {et~al.}(2011){Camargo}, {Bonatto}, \&
  {Bica}}]{2011MNRAS.416.1522C}
{Camargo}, D., {Bonatto}, C., \& {Bica}, E. 2011, \mnras, 416, 1522,
  \dodoi{10.1111/j.1365-2966.2011.19150.x}

\bibitem[{{Cao} {et~al.}(2023){Cao}, {Aharonian}, {An}, {Axikegu}, {Bai},
  {Bao}, {Bastieri}, {Bi}, {Bi}, {Cai}, {Cao}, {Cao}, {Cao}, {Chang}, {Chang},
  {Chen}, {Chen}, {Chen}, {Chen}, {Chen}, {Chen}, {Chen}, {Chen}, {Chen},
  {Chen}, {Chen}, {Chen}, {Cheng}, {Cheng}, {Cui}, {Cui}, {Cui}, {Cui}, {Dai},
  {Dai}, {Dai}, {Danzengluobu}, {della Volpe}, {Dong}, {Duan}, {Fan}, {Fan},
  {Fang}, {Fang}, {Feng}, {Feng}, {Feng}, {Feng}, {Feng}, {Gabici}, {Gao},
  {Gao}, {Gao}, {Gao}, {Gao}, {Gao}, {Ge}, {Geng}, {Giacinti}, {Gong}, {Gou},
  {Gu}, {Guo}, {Guo}, {Guo}, {Guo}, {Han}, {He}, {He}, {He}, {He}, {He},
  {Heller}, {Hor}, {Hou}, {Hou}, {Hou}, {Hu}, {Hu}, {Hu}, {Huang}, {Huang},
  {Huang}, {Huang}, {Huang}, {Huang}, {Huang}, {Ji}, {Jia}, {Jia}, {Jiang},
  {Jiang}, {Jiang}, {Jin}, {Kang}, {Ke}, {Kuleshov}, {Kurinov}, {Li}, {Li},
  {Li}, {Li}, {Li}, {Li}, {Li}, {Li}, {Li}, {Li}, {Li}, {Li}, {Li}, {Li}, {Li},
  {Li}, {Li}, {Li}, {Li}, {Liang}, {Liang}, {Lin}, {Liu}, {Liu}, {Liu}, {Liu},
  {Liu}, {Liu}, {Liu}, {Liu}, {Liu}, {Liu}, {Liu}, {Liu}, {Liu}, {Liu}, {Lu},
  {Luo}, {Lv}, {Ma}, {Ma}, {Ma}, {Mao}, {Min}, {Mitthumsiri}, {Mu}, {Nan},
  {Neronov}, {Ou}, {Pang}, {Pattarakijwanich}, {Pei}, {Qi}, {Qi}, {Qiao},
  {Qin}, {Ruffolo}, {S{\'a}iz}, {Semikoz}, {Shao}, {Shao}, {Shchegolev},
  {Sheng}, {Shu}, {Song}, {Stenkin}, {Stepanov}, {Su}, {Sun}, {Sun}, {Sun},
  {Tam}, {Tang}, {Tang}, {Tian}, {Wang}, {Wang}, {Wang}, {Wang}, {Wang},
  {Wang}, {Wang}, {Wang}, {Wang}, {Wang}, {Wang}, {Wang}, {Wang}, {Wang},
  {Wang}, {Wang}, {Wang}, {Wang}, {Wang}, {Wang}, {Wang}, {Wei}, {Wei}, {Wei},
  {Wen}, {Wu}, {Wu}, {Wu}, {Wu}, {Wu}, {Xi}, {Xia}, {Xia}, {Xiang}, {Xiao},
  {Xiao}, {Xin}, {Xin}, {Xing}, {Xiong}, {Xu}, {Xu}, {Xu}, {Xu}, {Xue}, {Yan},
  {Yan}, {Yan}, {Yang}, {Yang}, {Yang}, {Yang}, {Yang}, {Yang}, {Yang}, {Yang},
  {Yang}, {Yao}, {Yao}, {Ye}, {Yin}, {Yin}, {You}, {You}, {Yu}, {Yuan}, {Yue},
  {Zeng}, {Zeng}, {Zeng}, {Zha}, {Zhang}, {Zhang}, {Zhang}, {Zhang}, {Zhang},
  {Zhang}, {Zhang}, {Zhang}, {Zhang}, {Zhang}, {Zhang}, {Zhang}, {Zhang},
  {Zhang}, {Zhang}, {Zhang}, {Zhang}, {Zhang}, {Zhao}, {Zhao}, {Zhao}, {Zhao},
  {Zhao}, {Zheng}, {Zhou}, {Zhou}, {Zhou}, {Zhou}, {Zhou}, {Zhou}, {Zhou},
  {Zhu}, {Zhu}, {Zhu}, {Zhu}, \& {Zuo.}}]{2023arXiv230517030C}
{Cao}, Z., {Aharonian}, F., {An}, Q., {et~al.} 2023, arXiv e-prints,
  arXiv:2305.17030, \dodoi{10.48550/arXiv.2305.17030}

\bibitem[{{Cao} {et~al.}(2024){Cao}, {Aharonian}, {An}, {Axikegu}, {Bai},
  {Bao}, {Bastieri}, {Bi}, {Bi}, {Cai}, {Cao}, {Cao}, {Cao}, {Chang}, {Chang},
  {Chen}, {Chen}, {Chen}, {Chen}, {Chen}, {Chen}, {Chen}, {Chen}, {Chen},
  {Chen}, {Chen}, {Chen}, {Cheng}, {Cheng}, {Cui}, {Cui}, {Cui}, {Cui}, {Dai},
  {Dai}, {Dai}, {Danzengluobu}, {Della Volpe}, {Dong}, {Duan}, {Fan}, {Fan},
  {Fang}, {Fang}, {Feng}, {Feng}, {Feng}, {Feng}, {Feng}, {Gabici}, {Gao},
  {Gao}, {Gao}, {Gao}, {Gao}, {Gao}, {Ge}, {Geng}, {Giacinti}, {Gong}, {Gou},
  {Gu}, {Guo}, {Guo}, {Guo}, {Guo}, {Han}, {He}, {He}, {He}, {He}, {He},
  {Heller}, {Hor}, {Hou}, {Hou}, {Hou}, {Hu}, {Hu}, {Hu}, {Huang}, {Huang},
  {Huang}, {Huang}, {Huang}, {Huang}, {Huang}, {Ji}, {Jia}, {Jia}, {Jiang},
  {Jiang}, {Jiang}, {Jin}, {Kang}, {Ke}, {Kuleshov}, {Kurinov}, {Li}, {Li},
  {Li}, {Li}, {Li}, {Li}, {Li}, {Li}, {Li}, {Li}, {Li}, {Li}, {Li}, {Li}, {Li},
  {Li}, {Li}, {Li}, {Li}, {Liang}, {Liang}, {Lin}, {Liu}, {Liu}, {Liu}, {Liu},
  {Liu}, {Liu}, {Liu}, {Liu}, {Liu}, {Liu}, {Liu}, {Liu}, {Liu}, {Liu}, {Lu},
  {Luo}, {Lv}, {Ma}, {Ma}, {Ma}, {Mao}, {Min}, {Mitthumsiri}, {Mu}, {Nan},
  {Neronov}, {Ou}, {Pang}, {Pattarakijwanich}, {Pei}, {Qi}, {Qi}, {Qiao},
  {Qin}, {Ruffolo}, {S{\'a}iz}, {Semikoz}, {Shao}, {Shao}, {Shchegolev},
  {Sheng}, {Shu}, {Song}, {Stenkin}, {Stepanov}, {Su}, {Sun}, {Sun}, {Sun},
  {Tam}, {Tang}, {Tang}, {Tian}, {Wang}, {Wang}, {Wang}, {Wang}, {Wang},
  {Wang}, {Wang}, {Wang}, {Wang}, {Wang}, {Wang}, {Wang}, {Wang}, {Wang},
  {Wang}, {Wang}, {Wang}, {Wang}, {Wang}, {Wang}, {Wang}, {Wei}, {Wei}, {Wei},
  {Wen}, {Wu}, \& {Wu}}]{2024ApJS..271...25C}
---. 2024, \apjs, 271, 25, \dodoi{10.3847/1538-4365/acfd29}

\bibitem[{{Caprioli} \& {Spitkovsky}(2014)}]{2014ApJ...783...91C}
{Caprioli}, D., \& {Spitkovsky}, A. 2014, \apj, 783, 91,
  \dodoi{10.1088/0004-637X/783/2/91}

\bibitem[{{Castor} {et~al.}(1975){Castor}, {McCray}, \&
  {Weaver}}]{1975ApJ...200L.107C}
{Castor}, J., {McCray}, R., \& {Weaver}, R. 1975, \apjl, 200, L107,
  \dodoi{10.1086/181908}

\bibitem[{{Caswell}(1976)}]{1976MNRAS.177..601C}
{Caswell}, J.~L. 1976, \mnras, 177, 601, \dodoi{10.1093/mnras/177.3.601}

\bibitem[{{Cesarsky} \& {Montmerle}(1983)}]{1983SSRv...36..173C}
{Cesarsky}, C.~J., \& {Montmerle}, T. 1983, \ssr, 36, 173,
  \dodoi{10.1007/BF00167503}

\bibitem[{{Dame} {et~al.}(2001){Dame}, {Hartmann}, \&
  {Thaddeus}}]{2001ApJ...547..792D}
{Dame}, T.~M., {Hartmann}, D., \& {Thaddeus}, P. 2001, \apj, 547, 792,
  \dodoi{10.1086/318388}

\bibitem[{{Dewangan} \& {Anandarao}(2011)}]{2011MNRAS.414.1526D}
{Dewangan}, L.~K., \& {Anandarao}, B.~G. 2011, \mnras, 414, 1526,
  \dodoi{10.1111/j.1365-2966.2011.18487.x}

\bibitem[{{Evans} \& {Blair}(1981)}]{1981ApJ...246..394E}
{Evans}, II, N.~J., \& {Blair}, G.~N. 1981, \apj, 246, 394,
  \dodoi{10.1086/158937}

\bibitem[{{Gabici} {et~al.}(2009){Gabici}, {Aharonian}, \&
  {Casanova}}]{gabici2009}
{Gabici}, S., {Aharonian}, F.~A., \& {Casanova}, S. 2009, \mnras, 396, 1629,
  \dodoi{10.1111/j.1365-2966.2009.14832.x}

\bibitem[{{Gao} {et~al.}(2010){Gao}, {Reich}, {Han}, {Sun}, {Wielebinski},
  {Shi}, {Xiao}, {Reich}, {F{\"u}rst}, {Chen}, \& {Ma}}]{2010A&A...515A..64G}
{Gao}, X.~Y., {Reich}, W., {Han}, J.~L., {et~al.} 2010, \aap, 515, A64,
  \dodoi{10.1051/0004-6361/200913793}

\bibitem[{{Garmany} {et~al.}(1982){Garmany}, {Conti}, \&
  {Chiosi}}]{1982ApJ...263..777G}
{Garmany}, C.~D., {Conti}, P.~S., \& {Chiosi}, C. 1982, \apj, 263, 777,
  \dodoi{10.1086/160548}

\bibitem[{{Ge} {et~al.}(2022){Ge}, {Sun}, {Yang}, {Liang}, \&
  {Liang}}]{2022MNRAS.517.5121G}
{Ge}, T.-T., {Sun}, X.-N., {Yang}, R.-Z., {Liang}, Y.-F., \& {Liang}, E.-W.
  2022, \mnras, 517, 5121, \dodoi{10.1093/mnras/stac2885}

\bibitem[{{Giacinti} {et~al.}(2020){Giacinti}, {Mitchell}, {L{\'o}pez-Coto},
  {Joshi}, {Parsons}, \& {Hinton}}]{2020A&A...636A.113G}
{Giacinti}, G., {Mitchell}, A.~M.~W., {L{\'o}pez-Coto}, R., {et~al.} 2020,
  \aap, 636, A113, \dodoi{10.1051/0004-6361/201936505}

\bibitem[{{Grondin} {et~al.}(2013){Grondin}, {Romani}, {Lemoine-Goumard},
  {Guillemot}, {Harding}, \& {Reposeur}}]{2013ApJ...774..110G}
{Grondin}, M.~H., {Romani}, R.~W., {Lemoine-Goumard}, M., {et~al.} 2013, \apj,
  774, 110, \dodoi{10.1088/0004-637X/774/2/110}

\bibitem[{{H.~E.~S.~S. Collaboration} {et~al.}(2012){H.~E.~S.~S.
  Collaboration}, {Abramowski}, {Acero}, {Aharonian}, {Akhperjanian}, {Anton},
  {Balenderan}, {Balzer}, {Barnacka}, {Becherini}, {Becker}, {Bernl{\"o}hr},
  {Birsin}, {Biteau}, {Bochow}, {Boisson}, {Bolmont}, {Bordas}, {Brucker},
  {Brun}, {Brun}, {Bulik}, {B{\"u}sching}, {Carrigan}, {Casanova}, {Cerruti},
  {Chadwick}, {Charbonnier}, {Chaves}, {Cheesebrough}, {Cologna}, {Conrad},
  {Couturier}, {Dalton}, {Daniel}, {Davids}, {Degrange}, {Deil}, {Dickinson},
  {Djannati-Ata{\"\i}}, {Domainko}, {Drury}, {Dubus}, {Dutson}, {Dyks},
  {Dyrda}, {Egberts}, {Eger}, {Espigat}, {Fallon}, {Farnier}, {Fegan},
  {Feinstein}, {Fernandes}, {Fiasson}, {Fontaine}, {F{\"o}rster},
  {F{\"u}{\ss}ling}, {Gajdus}, {Gallant}, {Garrigoux}, {Gast}, {G{\'e}rard},
  {Giebels}, {Glicenstein}, {Gl{\"u}ck}, {G{\"o}ring}, {Grondin},
  {H{\"a}ffner}, {Hague}, {Hahn}, {Hampf}, {Harris}, {Hauser}, {Heinz},
  {Heinzelmann}, {Henri}, {Hermann}, {Hillert}, {Hinton}, {Hofmann},
  {Hofverberg}, {Holler}, {Horns}, {Jacholkowska}, {Jahn}, {Jamrozy}, {Jung},
  {Kastendieck}, {Katarzy{\'n}ski}, {Katz}, {Kaufmann}, {Kh{\'e}lifi},
  {Klochkov}, {Klu{\'z}niak}, {Kneiske}, {Komin}, {Kosack}, {Kossakowski},
  {Krayzel}, {Laffon}, {Lamanna}, {Lenain}, {Lennarz}, {Lohse}, {Lopatin},
  {Lu}, {Marandon}, {Marcowith}, {Masbou}, {Maurin}, {Maxted}, {Mayer},
  {McComb}, {Medina}, {M{\'e}hault}, {Menzler}, {Moderski}, {Mohamed},
  {Moulin}, {Naumann}, {Naumann-Godo}, {de Naurois}, {Nedbal}, {Nekrassov},
  {Nguyen}, {Nicholas}, {Niemiec}, {Nolan}, {Ohm}, {de O{\~n}a Wilhelmi},
  {Opitz}, {Ostrowski}, {Oya}, {Panter}, {Paz Arribas}, {Pekeur}, {Pelletier},
  {Perez}, {Petrucci}, {Peyaud}, {Pita}, {P{\"u}hlhofer}, {Punch},
  {Quirrenbach}, {Raue}, {Reimer}, {Reimer}, {Renaud}, {de los Reyes},
  {Rieger}, {Ripken}, {Rob}, {Rosier-Lees}, {Rowell}, {Rudak}, {Rulten},
  {Sahakian}, {Sanchez}, {Santangelo}, {Schlickeiser}, {Schulz}, {Schwanke},
  {Schwarzburg}, {Schwemmer}, {Sheidaei}, {Skilton}, {Sol}, {Spengler},
  {Stawarz}, {Steenkamp}, {Stegmann}, {Stinzing}, {Stycz}, {Sushch}, {Szostek},
  {Tavernet}, {Terrier}, {Tluczykont}, {Valerius}, {van Eldik}, {Vasileiadis},
  {Venter}, {Viana}, {Vincent}, {V{\"o}lk}, {Volpe}, {Vorobiov}, {Vorster},
  {Wagner}, {Ward}, {White}, {Wierzcholska}, {Zacharias}, {Zajczyk},
  {Zdziarski}, {Zech}, \& {Zechlin}}]{2012A&A...548A..46H}
{H.~E.~S.~S. Collaboration}, {Abramowski}, A., {Acero}, F., {et~al.} 2012,
  \aap, 548, A46, \dodoi{10.1051/0004-6361/201219814}

\bibitem[{{H.~E.~S.~S. Collaboration} {et~al.}(2018){H.~E.~S.~S.
  Collaboration}, {Abdalla}, {Abramowski}, {Aharonian}, {Ait Benkhali},
  {Ang{\"u}ner}, {Arakawa}, {Arrieta}, {Aubert}, {Backes}, {Balzer}, {Barnard},
  {Becherini}, {Becker Tjus}, {Berge}, {Bernhard}, {Bernl{\"o}hr}, {Blackwell},
  {B{\"o}ttcher}, {Boisson}, {Bolmont}, {Bonnefoy}, {Bordas}, {Bregeon},
  {Brun}, {Brun}, {Bryan}, {B{\"u}chele}, {Bulik}, {Capasso}, {Carrigan},
  {Caroff}, {Carosi}, {Casanova}, {Cerruti}, {Chakraborty}, {Chaves}, {Chen},
  {Chevalier}, {Colafrancesco}, {Condon}, {Conrad}, {Davids}, {Decock}, {Deil},
  {Devin}, {deWilt}, {Dirson}, {Djannati-Ata{\"\i}}, {Domainko}, {Donath},
  {Drury}, {Dutson}, {Dyks}, {Edwards}, {Egberts}, {Eger}, {Emery},
  {Ernenwein}, {Eschbach}, {Farnier}, {Fegan}, {Fernandes}, {Fiasson},
  {Fontaine}, {F{\"o}rster}, {Funk}, {F{\"u}{\ss}ling}, {Gabici}, {Gallant},
  {Garrigoux}, {Gast}, {Gat{\'e}}, {Giavitto}, {Giebels}, {Glawion},
  {Glicenstein}, {Gottschall}, {Grondin}, {Hahn}, {Haupt}, {Hawkes},
  {Heinzelmann}, {Henri}, {Hermann}, {Hinton}, {Hofmann}, {Hoischen}, {Holch},
  {Holler}, {Horns}, {Ivascenko}, {Iwasaki}, {Jacholkowska}, {Jamrozy},
  {Jankowsky}, {Jankowsky}, {Jingo}, {Jouvin}, {Jung-Richardt}, {Kastendieck},
  {Katarzy{\'n}ski}, {Katsuragawa}, {Katz}, {Kerszberg}, {Khangulyan},
  {Kh{\'e}lifi}, {King}, {Klepser}, {Klochkov}, {Klu{\'z}niak}, {Komin},
  {Kosack}, {Krakau}, {Kraus}, {Kr{\"u}ger}, {Laffon}, {Lamanna}, {Lau},
  {Lees}, {Lefaucheur}, {Lemi{\`e}re}, {Lemoine-Goumard}, {Lenain}, {Leser},
  {Lohse}, {Lorentz}, {Liu}, {L{\'o}pez-Coto}, {Lypova}, {Marandon},
  {Malyshev}, {Marcowith}, {Mariaud}, {Marx}, {Maurin}, {Maxted}, {Mayer},
  {Meintjes}, {Meyer}, {Mitchell}, {Moderski}, {Mohamed}, {Mohrmann},
  {Mor{\r{a}}}, {Moulin}, {Murach}, {Nakashima}, {de Naurois}, {Ndiyavala},
  {Niederwanger}, {Niemiec}, {Oakes}, {O'Brien}, {Odaka}, {Ohm}, {Ostrowski},
  {Oya}, {Padovani}, {Panter}, {Parsons}, {Paz Arribas}, {Pekeur}, {Pelletier},
  {Perennes}, {Petrucci}, {Peyaud}, {Piel}, {Pita}, {Poireau}, {Poon},
  {Prokhorov}, {Prokoph}, {P{\"u}hlhofer}, {Punch}, {Quirrenbach}, {Raab},
  {Rauth}, {Reimer}, {Reimer}, {Renaud}, {de los Reyes}, {Rieger}, {Rinchiuso},
  {Romoli}, {Rowell}, {Rudak}, {Rulten}, {Safi-Harb}, {Sahakian}, {Saito},
  {Sanchez}, {Santangelo}, {Sasaki}, {Schandri}, {Schlickeiser},
  {Sch{\"u}ssler}, {Schulz}, {Schwanke}, \& {Schwemmer}}]{2018A&A...612A...1H}
{H.~E.~S.~S. Collaboration}, {Abdalla}, H., {Abramowski}, A., {et~al.} 2018,
  \aap, 612, A1, \dodoi{10.1051/0004-6361/201732098}

\bibitem[{{Hanabata} {et~al.}(2014){Hanabata}, {Katagiri}, {Hewitt}, {Ballet},
  {Fukazawa}, {Fukui}, {Hayakawa}, {Lemoine-Goumard}, {Pedaletti}, {Strong},
  {Torres}, \& {Yamazaki}}]{hanabata2014detailed}
{Hanabata}, Y., {Katagiri}, H., {Hewitt}, J.~W., {et~al.} 2014, \apj, 786, 145,
  \dodoi{10.1088/0004-637X/786/2/145}

\bibitem[{{Hanaoka} {et~al.}(2019){Hanaoka}, {Kaneda}, {Suzuki}, {Kokusho},
  {Oyabu}, {Ishihara}, {Kohno}, {Furuta}, {Tsuchikawa}, \&
  {Saito}}]{2019PASJ...71....6H}
{Hanaoka}, M., {Kaneda}, H., {Suzuki}, T., {et~al.} 2019, \pasj, 71, 6,
  \dodoi{10.1093/pasj/psy126}

\bibitem[{{H{\"a}rer} {et~al.}(2023){H{\"a}rer}, {Reville}, {Hinton},
  {Mohrmann}, \& {Vieu}}]{2023A&A...671A...4H}
{H{\"a}rer}, L.~K., {Reville}, B., {Hinton}, J., {Mohrmann}, L., \& {Vieu}, T.
  2023, \aap, 671, A4, \dodoi{10.1051/0004-6361/202245444}

\bibitem[{{Harju} {et~al.}(1998){Harju}, {Lehtinen}, {Booth}, \&
  {Zinchenko}}]{1998A&AS..132..211H}
{Harju}, J., {Lehtinen}, K., {Booth}, R.~S., \& {Zinchenko}, I. 1998, \aaps,
  132, 211, \dodoi{10.1051/aas:1998448}

\bibitem[{{Helene}(1983)}]{helene1983}
{Helene}, O. 1983, Nuclear Instruments and Methods in Physics Research, 212,
  319, \dodoi{10.1016/0167-5087(83)90709-3}

\bibitem[{{Huber} {et~al.}(2013){Huber}, {Tchernin}, {Eckert}, {Farnier},
  {Manalaysay}, {Straumann}, \& {Walter}}]{2013A&A...560A..64H}
{Huber}, B., {Tchernin}, C., {Eckert}, D., {et~al.} 2013, \aap, 560, A64,
  \dodoi{10.1051/0004-6361/201321947}

\bibitem[{{Israel} \& {Felli}(1978)}]{1978A&A....63..325I}
{Israel}, F.~P., \& {Felli}, M. 1978, \aap, 63, 325

\bibitem[{{Kafexhiu} {et~al.}(2014){Kafexhiu}, {Aharonian}, {Taylor}, \&
  {Vila}}]{2014PhRvD..90l3014K}
{Kafexhiu}, E., {Aharonian}, F., {Taylor}, A.~M., \& {Vila}, G.~S. 2014, \prd,
  90, 123014, \dodoi{10.1103/PhysRevD.90.123014}

\bibitem[{{Kang} {et~al.}(2012){Kang}, {Koo}, \&
  {Salter}}]{2012AJ....143...75K}
{Kang}, J.-h., {Koo}, B.-C., \& {Salter}, C. 2012, \aj, 143, 75,
  \dodoi{10.1088/0004-6256/143/3/75}

\bibitem[{{Kerton} {et~al.}(2007){Kerton}, {Murphy}, \&
  {Patterson}}]{2007MNRAS.379..289K}
{Kerton}, C.~R., {Murphy}, J., \& {Patterson}, J. 2007, \mnras, 379, 289,
  \dodoi{10.1111/j.1365-2966.2007.11945.x}

\bibitem[{{Kharchenko} {et~al.}(2013){Kharchenko}, {Piskunov}, {Schilbach},
  {R{\"o}ser}, \& {Scholz}}]{2013A&A...558A..53K}
{Kharchenko}, N.~V., {Piskunov}, A.~E., {Schilbach}, E., {R{\"o}ser}, S., \&
  {Scholz}, R.~D. 2013, \aap, 558, A53, \dodoi{10.1051/0004-6361/201322302}

\bibitem[{{Kirsanova} {et~al.}(2008){Kirsanova}, {Sobolev}, {Thomasson},
  {Wiebe}, {Johansson}, \& {Seleznev}}]{2008MNRAS.388..729K}
{Kirsanova}, M.~S., {Sobolev}, A.~M., {Thomasson}, M., {et~al.} 2008, \mnras,
  388, 729, \dodoi{10.1111/j.1365-2966.2008.13429.x}

\bibitem[{{Lande} {et~al.}(2012){Lande}, {Ackermann}, {Allafort}, {Ballet},
  {Bechtol}, {Burnett}, {Cohen-Tanugi}, {Drlica-Wagner}, {Funk}, {Giordano},
  {Grondin}, {Kerr}, \& {Lemoine-Goumard}}]{2012ApJ...756....5L}
{Lande}, J., {Ackermann}, M., {Allafort}, A., {et~al.} 2012, \apj, 756, 5,
  \dodoi{10.1088/0004-637X/756/1/5}

\bibitem[{{Landecker} {et~al.}(2010){Landecker}, {Reich}, {Reid}, {Reich},
  {Wolleben}, {Kothes}, {Uyan{\i}ker}, {Gray}, {Del Rizzo}, {F{\"u}rst},
  {Taylor}, \& {Wielebinski}}]{2010A&A...520A..80L}
{Landecker}, T.~L., {Reich}, W., {Reid}, R.~I., {et~al.} 2010, \aap, 520, A80,
  \dodoi{10.1051/0004-6361/200913921}

\bibitem[{{Li} {et~al.}(2025){Li}, {Giacinti}, {Liu}, \&
  {Xing}}]{2025A&A...700A.143L}
{Li}, Y., {Giacinti}, G., {Liu}, S., \& {Xing}, Y. 2025, \aap, 700, A143,
  \dodoi{10.1051/0004-6361/202453342}

\bibitem[{{Li} {et~al.}(2024){Li}, {Liu}, \& {Giacinti}}]{2024A&A...689A.257L}
{Li}, Y., {Liu}, S., \& {Giacinti}, G. 2024, \aap, 689, A257,
  \dodoi{10.1051/0004-6361/202348873}

\bibitem[{{Li} {et~al.}(2023{\natexlab{a}}){Li}, {Liu}, \&
  {He}}]{2023ApJ...953..100L}
{Li}, Y., {Liu}, S., \& {He}, Y. 2023{\natexlab{a}}, \apj, 953, 100,
  \dodoi{10.3847/1538-4357/ace344}

\bibitem[{{Li} {et~al.}(2023{\natexlab{b}}){Li}, {Xin}, {Liu}, \&
  {He}}]{2023ApJ...945...21L}
{Li}, Y., {Xin}, Y., {Liu}, S., \& {He}, Y. 2023{\natexlab{b}}, \apj, 945, 21,
  \dodoi{10.3847/1538-4357/acb81d}

\bibitem[{{Liu} {et~al.}(2019){Liu}, {Yang}, {Sun}, {Aharonian}, \&
  {Chen}}]{2019ApJ...881...94L}
{Liu}, B., {Yang}, R.-z., {Sun}, X.-n., {Aharonian}, F., \& {Chen}, Y. 2019,
  \apj, 881, 94, \dodoi{10.3847/1538-4357/ab2df8}

\bibitem[{{Liu} {et~al.}(2020){Liu}, {Zeng}, {Xin}, \&
  {Zhu}}]{2020ApJ...897L..34L}
{Liu}, S., {Zeng}, H., {Xin}, Y., \& {Zhu}, H. 2020, \apjl, 897, L34,
  \dodoi{10.3847/2041-8213/ab9ff2}

\bibitem[{{L{\'o}pez-Coto} {et~al.}(2022){L{\'o}pez-Coto}, {de O{\~n}a
  Wilhelmi}, {Aharonian}, {Amato}, \& {Hinton}}]{2022NatAs...6..199L}
{L{\'o}pez-Coto}, R., {de O{\~n}a Wilhelmi}, E., {Aharonian}, F., {Amato}, E.,
  \& {Hinton}, J. 2022, Nature Astronomy, 6, 199,
  \dodoi{10.1038/s41550-021-01580-0}

\bibitem[{{L{\'o}pez-Coto} \& {Giacinti}(2018)}]{2018MNRAS.479.4526L}
{L{\'o}pez-Coto}, R., \& {Giacinti}, G. 2018, \mnras, 479, 4526,
  \dodoi{10.1093/mnras/sty1821}

\bibitem[{{Ma{\'\i}z-Apell{\'a}niz} {et~al.}(2004){Ma{\'\i}z-Apell{\'a}niz},
  {Walborn}, {Galu{\'e}}, \& {Wei}}]{2004ApJS..151..103M}
{Ma{\'\i}z-Apell{\'a}niz}, J., {Walborn}, N.~R., {Galu{\'e}}, H.~{\'A}., \&
  {Wei}, L.~H. 2004, \apjs, 151, 103, \dodoi{10.1086/381380}

\bibitem[{{Mattox} {et~al.}(1996){Mattox}, {Bertsch}, {Chiang}, {Dingus},
  {Digel}, {Esposito}, {Fierro}, {Hartman}, {Hunter}, {Kanbach}, {Kniffen},
  {Lin}, {Macomb}, {Mayer-Hasselwander}, {Michelson}, {von Montigny},
  {Mukherjee}, {Nolan}, {Ramanamurthy}, {Schneid}, {Sreekumar}, {Thompson}, \&
  {Willis}}]{mattox1996likelihood}
{Mattox}, J.~R., {Bertsch}, D.~L., {Chiang}, J., {et~al.} 1996, \apj, 461, 396,
  \dodoi{10.1086/177068}

\bibitem[{{Montmerle}(1979)}]{1979ApJ...231...95M}
{Montmerle}, T. 1979, \apj, 231, 95, \dodoi{10.1086/157166}

\bibitem[{{Morlino} {et~al.}(2021){Morlino}, {Blasi}, {Peretti}, \&
  {Cristofari}}]{2021MNRAS.504.6096M}
{Morlino}, G., {Blasi}, P., {Peretti}, E., \& {Cristofari}, P. 2021, \mnras,
  504, 6096, \dodoi{10.1093/mnras/stab690}

\bibitem[{{Permyakova} {et~al.}(2025){Permyakova}, {Carraro}, {Seleznev},
  {Sobolev}, {Ladeyschikov}, \& {Kirsanova}}]{2025ApJ...979..162P}
{Permyakova}, T.~A., {Carraro}, G., {Seleznev}, A.~F., {et~al.} 2025, \apj,
  979, 162, \dodoi{10.3847/1538-4357/ad957d}

\bibitem[{{Porter} {et~al.}(2006){Porter}, {Moskalenko}, \&
  {Strong}}]{2006ApJ...648L..29P}
{Porter}, T.~A., {Moskalenko}, I.~V., \& {Strong}, A.~W. 2006, \apjl, 648, L29,
  \dodoi{10.1086/507770}

\bibitem[{{Porter} {et~al.}(2008){Porter}, {Moskalenko}, {Strong}, {Orlando},
  \& {Bouchet}}]{2008ApJ...682..400P}
{Porter}, T.~A., {Moskalenko}, I.~V., {Strong}, A.~W., {Orlando}, E., \&
  {Bouchet}, L. 2008, \apj, 682, 400, \dodoi{10.1086/589615}

\bibitem[{{Ptuskin} {et~al.}(2006){Ptuskin}, {Moskalenko}, {Jones}, {Strong},
  \& {Zirakashvili}}]{2006ApJ...642..902P}
{Ptuskin}, V.~S., {Moskalenko}, I.~V., {Jones}, F.~C., {Strong}, A.~W., \&
  {Zirakashvili}, V.~N. 2006, \apj, 642, 902, \dodoi{10.1086/501117}

\bibitem[{{Reid} {et~al.}(2009){Reid}, {Menten}, {Zheng}, {Brunthaler},
  {Moscadelli}, {Xu}, {Zhang}, {Sato}, {Honma}, {Hirota}, {Hachisuka}, {Choi},
  {Moellenbrock}, \& {Bartkiewicz}}]{2009ApJ...700..137R}
{Reid}, M.~J., {Menten}, K.~M., {Zheng}, X.~W., {et~al.} 2009, \apj, 700, 137,
  \dodoi{10.1088/0004-637X/700/1/137}

\bibitem[{{Reid} {et~al.}(2014){Reid}, {Menten}, {Brunthaler}, {Zheng}, {Dame},
  {Xu}, {Wu}, {Zhang}, {Sanna}, {Sato}, {Hachisuka}, {Choi}, {Immer},
  {Moscadelli}, {Rygl}, \& {Bartkiewicz}}]{2014ApJ...783..130R}
{Reid}, M.~J., {Menten}, K.~M., {Brunthaler}, A., {et~al.} 2014, \apj, 783,
  130, \dodoi{10.1088/0004-637X/783/2/130}

\bibitem[{{Roger} {et~al.}(1999){Roger}, {Costain}, {Landecker}, \&
  {Swerdlyk}}]{1999A&AS..137....7R}
{Roger}, R.~S., {Costain}, C.~H., {Landecker}, T.~L., \& {Swerdlyk}, C.~M.
  1999, \aaps, 137, 7, \dodoi{10.1051/aas:1999239}

\bibitem[{{Ryabukhina} {et~al.}(2022){Ryabukhina}, {Kirsanova}, {Henkel}, \&
  {Wiebe}}]{2022MNRAS.517.4669R}
{Ryabukhina}, O.~L., {Kirsanova}, M.~S., {Henkel}, C., \& {Wiebe}, D.~S. 2022,
  \mnras, 517, 4669, \dodoi{10.1093/mnras/stac2877}

\bibitem[{{Snowden} {et~al.}(1997){Snowden}, {Egger}, {Freyberg}, {McCammon},
  {Plucinsky}, {Sanders}, {Schmitt}, {Tr{\"u}mper}, \&
  {Voges}}]{1997ApJ...485..125S}
{Snowden}, S.~L., {Egger}, R., {Freyberg}, M.~J., {et~al.} 1997, \apj, 485,
  125, \dodoi{10.1086/304399}

\bibitem[{{Solin} {et~al.}(2012){Solin}, {Ukkonen}, \&
  {Haikala}}]{2012A&A...542A...3S}
{Solin}, O., {Ukkonen}, E., \& {Haikala}, L. 2012, \aap, 542, A3,
  \dodoi{10.1051/0004-6361/201118531}

\bibitem[{{Sun} {et~al.}(2020{\natexlab{a}}){Sun}, {Yang}, {Liang}, {Peng},
  {Zhang}, {Wang}, \& {Aharonian}}]{2020A&A...639A..80S}
{Sun}, X.-N., {Yang}, R.-Z., {Liang}, Y.-F., {et~al.} 2020{\natexlab{a}}, \aap,
  639, A80, \dodoi{10.1051/0004-6361/202037580}

\bibitem[{{Sun} {et~al.}(2020{\natexlab{b}}){Sun}, {Yang}, \&
  {Wang}}]{2020MNRAS.494.3405S}
{Sun}, X.-N., {Yang}, R.-Z., \& {Wang}, X.-Y. 2020{\natexlab{b}}, \mnras, 494,
  3405, \dodoi{10.1093/mnras/staa947}

\bibitem[{{Thoudam} \& {H{\"o}randel}(2012)}]{2012MNRAS.419..624T}
{Thoudam}, S., \& {H{\"o}randel}, J.~R. 2012, \mnras, 419, 624,
  \dodoi{10.1111/j.1365-2966.2011.19724.x}

\bibitem[{{Uchiyama} {et~al.}(2012){Uchiyama}, {Funk}, {Katagiri}, {Katsuta},
  {Lemoine-Goumard}, {Tajima}, {Tanaka}, \& {Torres}}]{uchiyama2012fermi}
{Uchiyama}, Y., {Funk}, S., {Katagiri}, H., {et~al.} 2012, \apjl, 749, L35,
  \dodoi{10.1088/2041-8205/749/2/L35}

\bibitem[{{Wood} {et~al.}(2017){Wood}, {Caputo}, {Charles}, {Di Mauro},
  {Magill}, {Perkins}, \& {Fermi-LAT Collaboration}}]{2017ICRC...35..824W}
{Wood}, M., {Caputo}, R., {Charles}, E., {et~al.} 2017, in International Cosmic
  Ray Conference, Vol. 301, 35th International Cosmic Ray Conference
  (ICRC2017), 824.
\newblock \doarXiv{1707.09551}

\bibitem[{{Yang} {et~al.}(2018){Yang}, {de O{\~n}a Wilhelmi}, \&
  {Aharonian}}]{2018A&A...611A..77Y}
{Yang}, R.-z., {de O{\~n}a Wilhelmi}, E., \& {Aharonian}, F. 2018, \aap, 611,
  A77, \dodoi{10.1051/0004-6361/201732045}

\bibitem[{{Zabalza}(2015)}]{zabalza2015naima}
{Zabalza}, V. 2015, in International Cosmic Ray Conference, Vol.~34, 34th
  International Cosmic Ray Conference (ICRC2015), 922.
\newblock \doarXiv{1509.03319}

\bibitem[{{Zeng} {et~al.}(2019){Zeng}, {Xin}, \& {Liu}}]{2019ApJ...874...50Z}
{Zeng}, H., {Xin}, Y., \& {Liu}, S. 2019, \apj, 874, 50,
  \dodoi{10.3847/1538-4357/aaf392}

\end{thebibliography}
\bibliographystyle{aasjournal.bst}

\section*{\textbf{Appendix: SNR-MC directly interactions for hadronci scenario}}

In this section, we explore an alternative scenario in which SrcA originates from SNR shock–cloud interactions, while SrcB results from escaped CRs illuminating nearby high-density gas. This interpretation differs from that in Section~\ref{sec:4.1}, where both SrcA and SrcB were assumed to be produced and dominated by escaped CR contributions. In the present model, the spectrum of SrcA is described by a simple power law with an index of $\sim 2.4$ (consistent with the injection spectrum index) and a cutoff around 200 TeV. SrcB is still attributed to escaped CRs interacting with nearby gas under Kraichnan turbulence with different $\chi$ value tested, while all other parameters are kept the same. The only differences from Section~\ref{sec:4.1} are that the injected spectral index is set to 2.4 instead of 2.0, and the distance parameter $r_{\rm s}$ is slightly increased from 50 pc to 60 pc.

\renewcommand{\thefigure}{A}
\setcounter{figure}{0}
\begin{figure*}[htbp]
    \centering
    \includegraphics[trim={0 0.cm 0 0}, clip,width=0.42\textwidth]{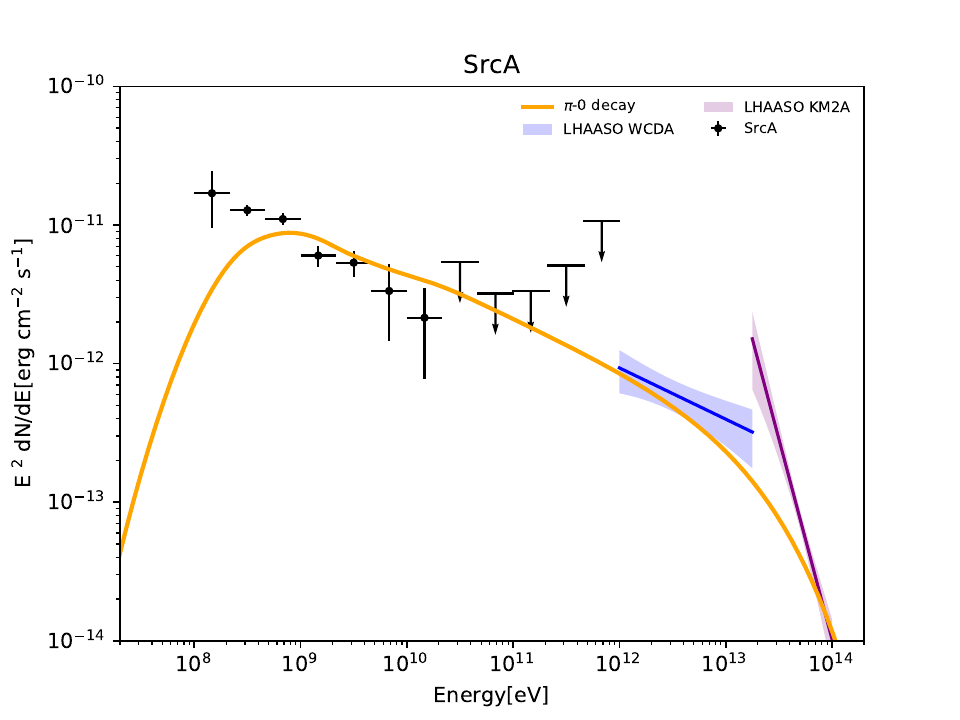}
    \includegraphics[trim={0 0.cm 0 0}, clip,width=0.42\textwidth]{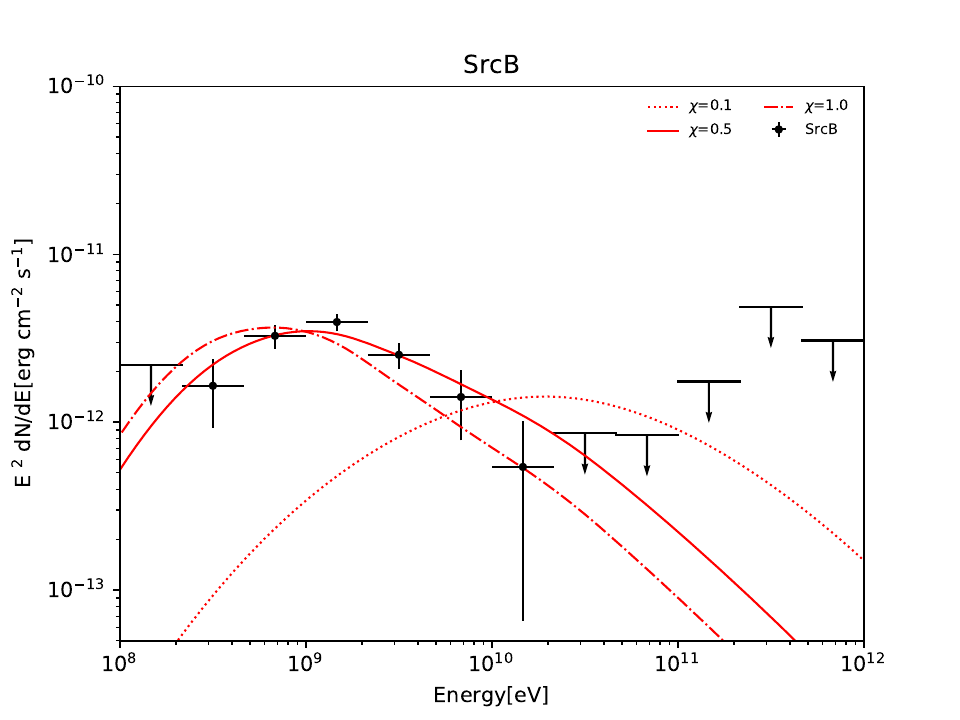}\\
    \caption{Hadronic modeling for SrcA and SrcB, respectively.}
   \label{fig:A}
\end{figure*}

\end{document}